\documentclass[aps,pre,twocolumn,superscriptaddress,groupedaddress]{revtex4}  % for review and submission

\usepackage{graphicx}  % needed for figures
\usepackage{placeins}
\usepackage{url}                  %this package should fix any errors with URLs in refs.
\usepackage{dcolumn}       % needed for some tables
\usepackage{bm}                  % for math
\usepackage{amssymb}       % for math
\usepackage{mathtools}     % for math
\usepackage{booktabs}
\usepackage{pgf}                   %plots from matplotlib in TeX style
\usepackage{tikz}                   %draw graphics
\usepackage{lipsum}

\newcommand{\LTmean}[1]{\langle #1 \rangle_{x,t}}               % line time average
\newcommand{\Vmean}[1]{\langle  #1 \rangle_{x,z}}               % volume average
\newcommand{\Tmean}[1]{\langle  #1 \rangle_{t}}                 % time average
\newcommand{\Lmean}[1]{\langle  #1 \rangle_{x}}                 % line (x) average

\newcommand{\partialD}[2]{\frac{\partial #1}{\partial #2}}      % partial derivative
  {\end{minipage}\vskip\textfloatsep\noindent}
  
\renewcommand{\arraystretch}{1.2}

\begin{document}

\title{Generalizability of reservoir computing for flux-driven two-dimensional convection}

\author{Florian Heyder}
\affiliation{Institut f\"ur Thermo- und Fluiddynamik, Technische Universit\"at Ilmenau, Postfach 100565, D-98684 Ilmenau, Germany}

\author{Juan Pedro Mellado}
\affiliation{Meteorologisches Institut, Universit\"at Hamburg, Bundesstra\ss e 55, D-20146 Hamburg, Germany}

\author{J\"org Schumacher}
\affiliation{Institut f\"ur Thermo- und Fluiddynamik, Technische Universit\"at Ilmenau, Postfach 100565, D-98684 Ilmenau, Germany}
\affiliation{Tandon School of Engineering, New York University, New York City, NY 11201, USA}

\date{\today}

\begin{abstract}
We explore the generalization properties of an echo state network applied as a reduced dynamical model to predict flux-driven two-dimensional turbulent convection. To this end, we consider a convection domain at fixed height with a variable ratio of buoyancy fluxes at the top and bottom boundaries, which break the top-down symmetry in comparison to the standard Rayleigh-B\'enard case thus leading to highly asymmetric mean and fluctuation profiles across the layer. Our direct numerical simulation model describes a convective boundary layer in a simple way. The data are used to train and test a recurrent neural network in the form of an echo state network. The input to the echo state networks is obtained in two different ways, either by a proper orthogonal decomposition or by a convolutional autoencoder. In both cases, the echo state network reproduces the turbulence dynamics and the statistical properties of the buoyancy flux, and is able to model unseen data records with different flux ratios.
\end{abstract}

\maketitle
%---------------------------------------------------------------------
%                     Introduction
%---------------------------------------------------------------------
\section{Introduction}
{\label{sec:introduction}}

Machine learning (ML) methods are known for their exceptional capabilities in the classification of comprehensive data records and data-driven modelling. In fluid mechanics, ML thus found its way into the analysis and control of turbulent flows~\cite{Kutz2017,Brenner2019,Duraisamy2019,Brunton2020,Pandey2020a,BeckKurzGAMM2021}. Applications cover now a broad spectrum of problems, such as the subgrid scale modeling in Reynolds-averaged Navier-Stokes equations \cite{Ling2016} or large-eddy simulations \cite{BeckJCP2019,KoumutsakosNatMachine}, the exploration of inertial manifolds in the phase space of the systems \cite{GrahamPRE2020} and the control of their spatio-temporal dynamics \cite{ZengGraham2021}, or the reconstruction and generation of partially missing turbulent data by deep neural networks \cite{Buzzicotti2021,Fukami2021}. The aim of such ML applications is usually to reduce the computational cost which comes with solving the governing equations of motion in direct numerical simulations or analysing high-resolution experimental data \cite{SchroederNatMachine2021}. Among other fluid flows, turbulent convection has been chosen as a prominent application case. We mention the classification of convective heat flux patterns \cite{Fonda2019}, spectral nudging methods to reconstruct the flow fields from temperature measurements \cite{AgasthyaPoF2022}, the application of reinforcement learning to control the heat transport in a Rayleigh-Be\'enard cell~\cite{Beintema2020}, or the prediction and reconstruction of turbulent dry and moist convection flows by recurrent neural networks \cite{Pandey2020,Heyder2021,Pandey2022,Valori2022}. Convection plays a prominent role in geophysical flows \cite{Wyngaard:2010,Mellado2017}. Machine learning is then applied for parameterizations of unresolved convection processes in the oceans \cite{Zanna2020} and atmosphere \cite{Wang2021,Yuval2021}. Given the strong variability in the environmental conditions in atmospheric flows, particular interest lies on the development of robust ML methods which can model configurations that are different from the training cases with respect to the parameter setting \cite{Goodfellow-et-al-2016}. This particular point sets the stage for the present work which consists of two major parts.

In the first part, we discuss a two-dimensional Rayleigh-B\'enard convection (RBC) model \cite{Chilla2012} that is driven by heat or buoyancy fluxes from the top and bottom. In our direct numerical simulations (DNS), we consider a convective cell with constant height $H$ in which bottom and top fluxes are chosen such that the cell as a whole gets differently strongly heated from the bottom and the top. This configuration can be understood as a simplified model of a convective boundary layer (CBL) in cloud-free and shear-free conditions. In this model, we retain the entrainment of fluid from the free troposphere into the turbulent region by prescribed top flux into the convective domain as well as heating from below from the heated ground. A difference between this model and a real CBL is that $H$ in our model remains constant, whereas the height of the atmospheric layer increases slowly with increasing time in reality, i.e., the CBL grows into the free troposphere \cite{adrian1986,Zilitinkevich:1991,Sorbjan:1996,fodor2019}. Given this setup, the top-down symmetry of a standard RBC flow will be broken; highly asymmetric mean profiles of buoyancy, convective buoyancy flux and velocity fluctuations follow. This clearly challenges the reproduction of statistical properties by the ML algorithm. 

In the second part, we use the DNS data as a training data base to study the performance of {\em dynamical} reduced-order models of turbulent convection based on recurrent neural network architectures, i.e.,  neural networks with a short-term memory. These ML algorithms will then be applied to data that have a different ratio of boundary fluxes as the training configuration, i.e, have a different set of system parameters. Our study thus addresses one important point of supervised ML algorithms, namely how well do they perform with respect to unseen data with changed system parameters -- known as the generalization property or generalizability \cite{Goodfellow-et-al-2016}. More specifically, we apply echo state networks (ESN) which are one implementation of reservoir computing \citep{JaegerHaas2004,LukoseviciusJaeger2012}. The ESN approach has found interest recently in inferring states of a nonlinear dynamical system. Applications concerned the R\"ossler and Lorenz 63 systems ~\citep{lu_reservoir_2017,pathak_using_2017}, the Lorenz 96 model \cite{Vlachas2020}, and Galerkin models of plane shear flows \cite{Doan2021}. Moreover, hybrid models which combine both data driven (ESN) and knowledge based methods, i.e. solving the mathematical equations, have already been proposed~\citep{pathak2018Chaos,Wikner2020Chaos} and tested in terms of a global atmospheric forecast model~\citep{Arcomano2020}. Further, reservoir computing techniques, due to their computationally inexpensive training routine, could serve as a substitute for conventional parameterization schemes. The performance of ESNs in two-dimensional dry and moist turbulent Rayleigh-B\'enard convection have already shown great promise, as low-order statistics of buoyancy and liquid water fluxes are successfully reproduced~\citep{Pandey2020,Heyder2021,Pandey2022}. Here, we want to apply this framework to a case that is a bit closer to real atmospheric flows than standard RBC.

Even two-dimensional DNS data records are still too large to be directly processed by the ESN. Thus, a data reduction step is required. We suggest two methods here, (1) the proper orthogonal decomposition (POD) and (2) the convolutional autoencoder (CAE) \cite{Sirovich1987,Bailon2011,Baldi2012,Gonzalez2018,Pandey2022}. As a consequence, the present ML algorithm is a combination of two building blocks, the encoder-decoder module and the dynamical core in the form of an ESN which advances the convection flow in time in the low-dimensional latent space. It is found that, despite smaller differences in flux statistics and reconstruction of the fields, both models perform well. We thus investigate (1) the performance of two data reduction methods, namely POD and CAE, on turbulent convection data, (2) the generalization capability of the ESN to data with a different heat flux ratio (which will be defined in the next section), and (3) the combined application of data reduction and reservoir computing to flux-driven highly asymmetric Rayleigh-B\'enard convection flow.

The outline of the manuscript is as follows. In section 2, we describe the two-dimensional convection model and define all parameters, in particular, the ratio of the buoyancy fluxes at the top and bottom boundaries, $\beta$, which is the major control parameter. Section 3 introduces POD, CAE, and ESN. We provide details on the training and the generalization performance of the ESN. We summarize our results and give a brief outlook in the final section 4. Technical details on ML are listed in the appendices.

%---------------------------------------------------------------------
%                     Physics
%---------------------------------------------------------------------
\section{Flux-driven convection model}
{\label{sec:boussinesq_model}}
\subsection{Governing equations and model parameters}

\begin{figure}
    \centering
    \includegraphics[width = .45\textwidth]{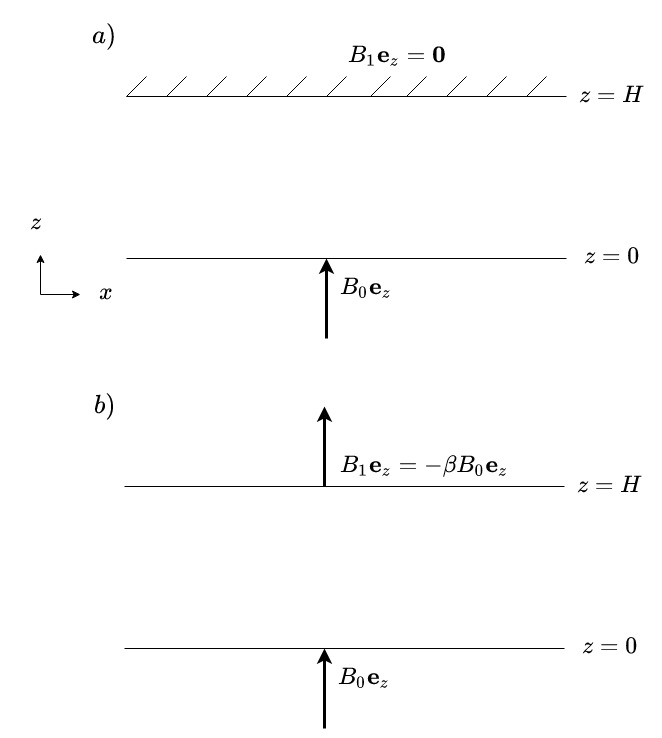}
    \caption{Scheme of the two-dimensional Rayleigh-B\'enard setup with constant buoyancy flux boundary conditions. The bottom of the cell is heated by the incoming flux $B_0$. We explore the effect of asymmetric boundary conditions by imposing a different flux $B_1 = -\beta B_0$ ($\beta > 0$) at the top. The values $\beta \in \{0.1,0.2,0.3\}$ are representative values for convective boundary layers. Here we consider (a) Adiabatic top ($\beta = 0$) and (b) warming flux at the top ($\beta > 0$).}
    \label{fig:flux_concept}
\end{figure}

We use the Boussinesq approximation to the two-dimensional Navier-Stokes equations. For convenience, we formulate the problem in terms of the buoyancy field $b$ which is given by 
\begin{equation}
b(x,z,t)\equiv\alpha g T(x,z,t)\,,
\end{equation}
where $\alpha, g$ and $T$ are the thermal expansion coefficient, the acceleration due to gravity and the temperature field, respectively. We consider a cell of height $H$ and length $L$ (see figure \ref{fig:flux_concept}). In the vertical direction, we consider no-slip boundary conditions for the velocity and constant-flux boundary conditions for the buoyancy. We impose the fluxes $B_0$ and $B_1$ at the bottom and top respectively. In the horizontal direction, we consider periodic boundary conditions.

The resulting evolution equations are given by
%---------------------------
\begin{align}
    \label{eq:gov_incomp}
   \partialD{u_x}{x}+\partialD{u_z}{z} & = 0\\
    \label{eq:gov_ux}
   \partialD{u_x}{t} +u_x\partialD{u_x}{x}+u_z\partialD{u_x}{z} & = -\partialD{p}{x}\nonumber\\ &+
   \nu\left(\frac{\partial^2 u_x}{\partial x^2}+\frac{\partial^2 u_x}{\partial z^2}\right)\\
   \label{eq:gov_uy}
   \partialD{u_z}{t} +u_x\partialD{u_z}{x}+u_z\partialD{u_z}{z} & = -\partialD{p}{z}\nonumber\\ &+ \nu\left(\frac{\partial^2 u_z}{\partial x^2}+\frac{\partial^2 u_z}{\partial z^2}\right)
   +b\\
   \label{eq:gov_b}
   \partialD{b}{t} + u_x\partialD{b}{x}+u_z\partialD{b}{z} &= \kappa\left(\frac{\partial^2 b}{\partial x^2}+\frac{\partial^2 b}{\partial z^2}\right)
\end{align}
%---------------------------
In this equations, $u_x$ and $u_z$ are the horizontal and vertical components of the velocity, $p$ is the modified pressure divided by the density, $\nu$ is the kinematic viscosity and $\kappa$ is the molecular diffusivity. The boundary conditions are $u_x=0$ and $u_z=0$ at $z=0$ and $z=1$, together with
%---------------------------
\begin{align}
    &\partialD{b}{z}(x,z=0,t)=-B_0/\kappa\\
    &\partialD{b}{z}(x,z=1,t)=-B_1/\kappa
\end{align}
%---------------------------
For the sake of generality, we will present the analysis in a non-dimensional form. Choosing $H$ and $B_0$ as reference scales, one finds the following characteristic scales: the convective velocity $(B_0H)^{1/3}$, the convective time $(H^2/B_0)^{1/3}$, and the convective buoyancy $(B_0^2/H)^{1/3}$ \citep{Deardorff:1970}. The resulting four controlling parameters are the aspect ratio $\Gamma=L/H$, the Prandtl number
%---------------------------
\begin{equation}
\textrm{Pr} = \frac{\nu}{\kappa} \;,
\end{equation}
%---------------------------
the convective Rayleigh number 
%---------------------------
\begin{equation}
\textrm{Ra}_c  = \frac{B_0 H^4}{\nu \kappa^2} \;,   
\end{equation}
and the buoyancy-flux ratio
%---------------------------
\begin{equation}
	\label{eq:top_flux_beta}
    \beta=-\frac{B_1}{B_0}.
\end{equation}
%---------------------------
As further explained below, we are interested in the cases $B_0>$ and $B_1<0$ and hence $\beta>0$, i.e., the fluid is heated from the bottom and from the top.

The buoyancy difference 
%---------------------------
\begin{equation}
\label{eq:buoy_diff}
    \Delta b = \Lmean{b}(z=0,t) -\Lmean{b}(z=1,t)
\end{equation}
%---------------------------
between the two plates is a dependent variable for configurations with constant-flux boundary conditions and needs to be diagnosed from experimental or simulation data (angle brackets indicate an averaging operation and the subscript indicates the variable with respect to which the averaging operation is performed, in this case, the horizontal coordinate $x$). Therefore, the Dirichlet Rayleigh number
%---------------------------
\begin{equation}
    \textrm{Ra}_f = \frac{\Delta b H^3}{\nu\kappa}
\end{equation}
%---------------------------
is a diagnostic variable as well. From eqns. \eqref{eq:gov_incomp}--\eqref{eq:gov_b} one can derive the vertical buoyancy profile, up to a constant, for the purely conductive case to be
%---------------------------
\begin{align}
\label{eq:buoy_conduct}
    b_\textrm{cond} =& \frac{B_0H}{\kappa}\frac{(z/H-1)^2+\beta (z/H)^2}{2}\\ &+\frac{B_0}{H}(1+\beta)t+\textrm{constant}.
\end{align}
%---------------------------
For $\beta=-1$, we recover the steady, linear solution that corresponds to the problem with Dirichlet boundary conditions. In the Neumann case, the profile of $b$ has a parabolic shape and it grows linearly in time (given that we heat from below and from above). Nonetheless, it is quasi-steady in the sense that the shape of the profile remains constant in time. The buoyancy difference between bottom and top plate for this case is 
%---------------------------
\begin{equation}
	\label{eq:buoy_diff_conduct}
    \Delta b_\textrm{cond} = \frac{B_0H}{\kappa}\frac{1-\beta}{2}.
\end{equation}
%---------------------------

%------------------------------------------------------
%-------------------------------------------------------
\subsection{Direct numerical simulations}
We fix the Prandtl and convective Rayleigh number to $\textrm{Pr} = 1$ and $\textrm{Ra}_c=3\cdot 10^8$ and consider extended layers with an aspect ratio $\Gamma=L/H=24$. The control parameter that we vary is the flux-ratio parameter $\beta$ defined by eq.~\eqref{eq:top_flux_beta}. In the atmospheric CBL over land, one typically finds the conditions $B_0>0$ and $B_1<0$, which represent the surface warming and the entrainment warming of the CBL, respectively. Hence, we are interested in the case $\beta > 0$. Typical atmospheric conditions correspond to the range $\beta\approx 0.1$ to $0.3$ \citep{Stull:1988,Wyngaard:2010}. As $\beta$ increases, the upper region of the convective cell increasingly stabilizes (positive mean buoyancy gradient, see also later in figure~\ref{fig:vertical_profiles}), and preliminary simulations (not shown) indicate that the dynamics strongly change for values $\beta\approx 0.4$. Therefore, we consider the cases $\beta\in\{0.0,0.1,0.2,0.3\}$ in our CBL model (see also figure \ref{fig:flux_concept}). The case $\beta=0$ corresponds to an upper adiabatic wall. This case is considered as a first step to understand the effect of asymmetries in the boundary conditions in the results obtained from Rayleigh-B\'enard convection with constant-buoyancy boundaries. 

The Boussinesq equations~\eqref{eq:gov_incomp} -- \eqref{eq:gov_b} are discretized by a high-order spectral-like compact finite difference method. The time evolution is treated by a low-storage fourth-order Runge-Kutta scheme. The pressure-Poisson equation is solved with a Fourier decomposition in the horizontal planes and a factorization of the resulting difference equations in the vertical direction. More details on the numerical method can be found in Mellado and Ansorge \cite{Mellado2012}. The software used to perform the simulations is freely available at \url{https://github.com/turbulencia/tlab}.

The grid size is $N_x\times N_z = 2400\times 150$. The horizontal grid spacing is uniform. The vertical grid spacing follows a hyperbolic tangent profile: it is equal within $1.2\%$ to the horizontal grid spacing in the center of the convection cell, and diminishes by a factor of 2.5 next to the wall. The time steps are in the range $\Delta t\approx 0.0012-0.0016\,(H^2/B_0)^{1/3}$, the specific value depending on the simulation. They are defined to obtain data exactly every $0.25$ free-fall times (definition follows) and satisfy the stability constraints of the numerical algorithm described in the previous paragraph. Since the free-fall time is a derived variable in the case of constant-flux boundaries considered in this study, preliminary simulations were performed to obtain the free-fall time in each case, and we repeated the simulations with the appropriate $\Delta t$. Table \ref{tab:1} summarizes important parameters of the four simulation runs. 

\begin{table}
    \renewcommand{\arraystretch}{1.5}
    \centering
    \begin{tabular}{ccccc}
         \hline\hline
         $\beta$ & $\langle\Delta b\rangle_t $ & $\langle T_f\rangle_t$& $\langle\mathrm{Nu}_f\rangle_t$ & $\langle\mathrm{Ra}_f\rangle_t$ \\
         \hline
         0.0 & $22.0\pm 0.4$ & 0.21 & $15.2\pm 0.3$ & $9.9\times 10^6$ \\
         0.1 & $19.0\pm 0.4$ & 0.23 & $15.9\pm 0.3$ & $8.5\times 10^6$ \\
         0.2 & $15.4\pm 0.3$ & 0.26 & $17.4\pm 0.4$ & $6.9\times 10^6$ \\
         0.3 & $10.7\pm 0.5$ & 0.31 & $21.9\pm 0.9$ & $4.8\times 10^6$ \\
         \hline\hline
    \end{tabular}
    \caption{Simulation parameters. The time average has been calculated over the last 500 free-fall times. The buoyancy and time values in the second and third columns are given in units of convective buoyancy $(B_0^2/H)^{1/3}$ and convective time $(H^2/B_0)^{1/3}$, respectively}
    \label{tab:1}
\end{table}
\subsection{Cellular convection patterns and vertical profiles at different flux ratios}
\begin{figure}
    \centering
    \includegraphics[width = 0.4\textwidth]{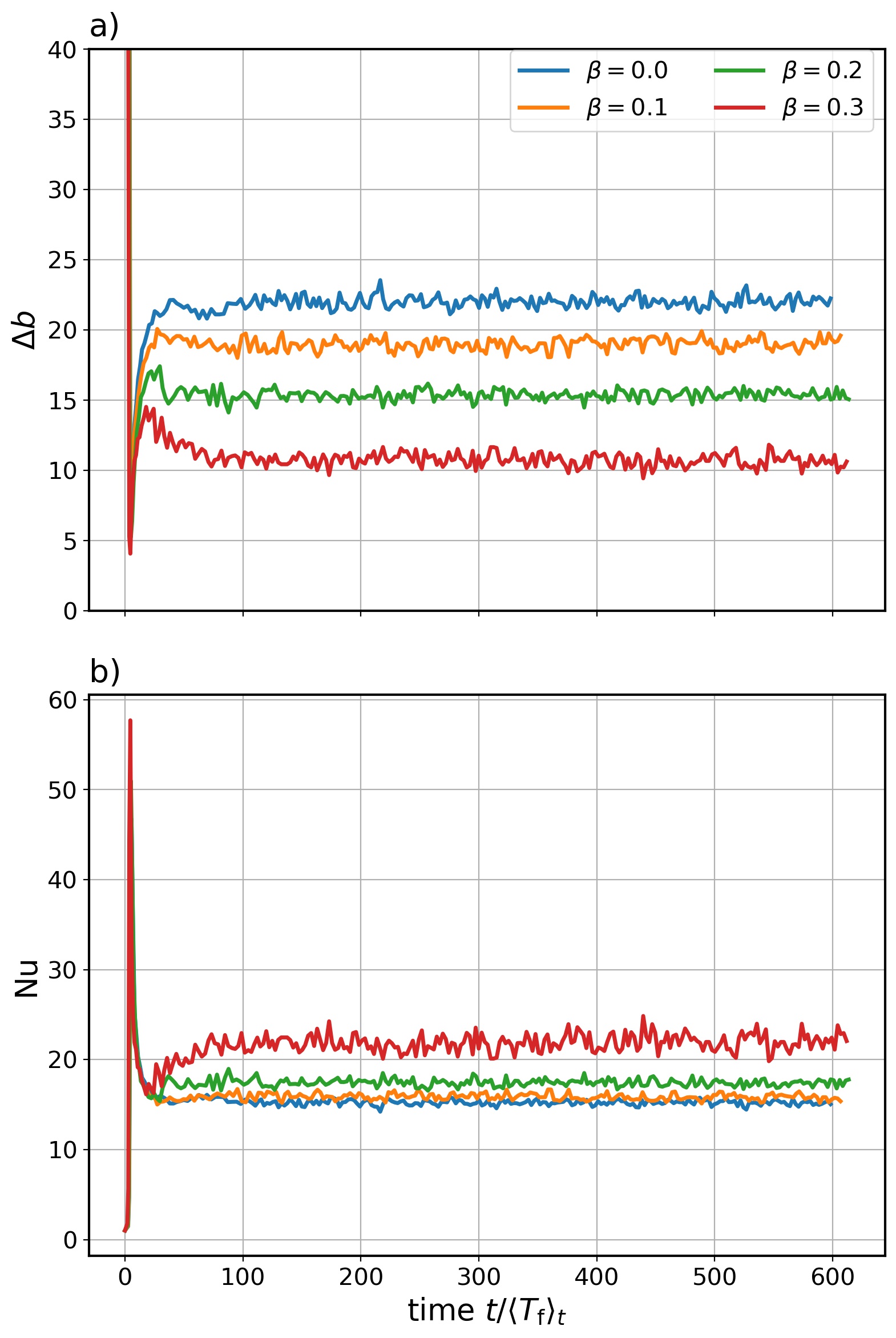}
    \caption{Temporal variation of (a) the buoyancy difference $\Delta b$, see eq.~\eqref{eq:buoy_diff}, and (b) the Nusselt number Nu, see eq.~\eqref{eq:nusselt}. Both quantities become statistically stationary after an initial transient. Note that both time axis are normalized by the time mean of the free fall time $T_f = \sqrt{H/\Delta b}$ in the statistical stationary regime.}
    \label{fig:diagnostics_temporal}
\end{figure}

Integrating the evolution equation for $b$ yields that the volume averaged buoyancy $\Vmean{b}$ increases as
\begin{equation}
    \Vmean{b} = \frac{B_0}{H}(1+\beta)t.
\end{equation}
Hence, in the turbulent case, the fluid warms linearly with increasing time as in the conduction case. The mean vertical profile, however, is different to the pure conduction profile and, as mentioned above, a major dependent variable is the buoyancy difference $\Delta b$ across the cell. After an initial transient, this quantity becomes statistically stationary, as can be seen in figure \ref{fig:diagnostics_temporal}(a) for all four simulations.
%------------------------------------------------------
%------------------------------------------------------
The free fall time $T_f = \sqrt{H/\Delta b}$ and free fall velocity $U_f = \sqrt{H\Delta b}$ can be computed and used as scales for better comparison to the more common case of Rayleigh-B\'enard convection with constant-buoyancy boundaries. Moreover, we can express the buoyancy difference in terms of a Nusselt number
%---------------------------
\begin{equation}
\label{eq:nusselt}
    \textrm{Nu} = \frac{\Delta b_\textrm{cond}}{\Delta b} = \frac{1-\beta}{2}\frac{B_0H}{\kappa\Delta b},
\end{equation}
%---------------------------
defined here as the ratio between the buoyancy difference in the purely conductive case $\Delta b_\textrm{cond}$ (see eq.~\eqref{eq:buoy_diff_conduct}), and the fully convective case, i.e. $\Delta b$. For $\beta=-1$, we again recover the functional relationship corresponding to Rayleigh-B\'enard convection with constant-buoyancy boundaries. The relaxation to a statistically stationary state for the buoyancy difference and the Nusselt number are demonstrated in figure \ref{fig:diagnostics_temporal} for all four cases. 

\begin{figure*}
\centering
    \includegraphics[width = 1.0\textwidth]{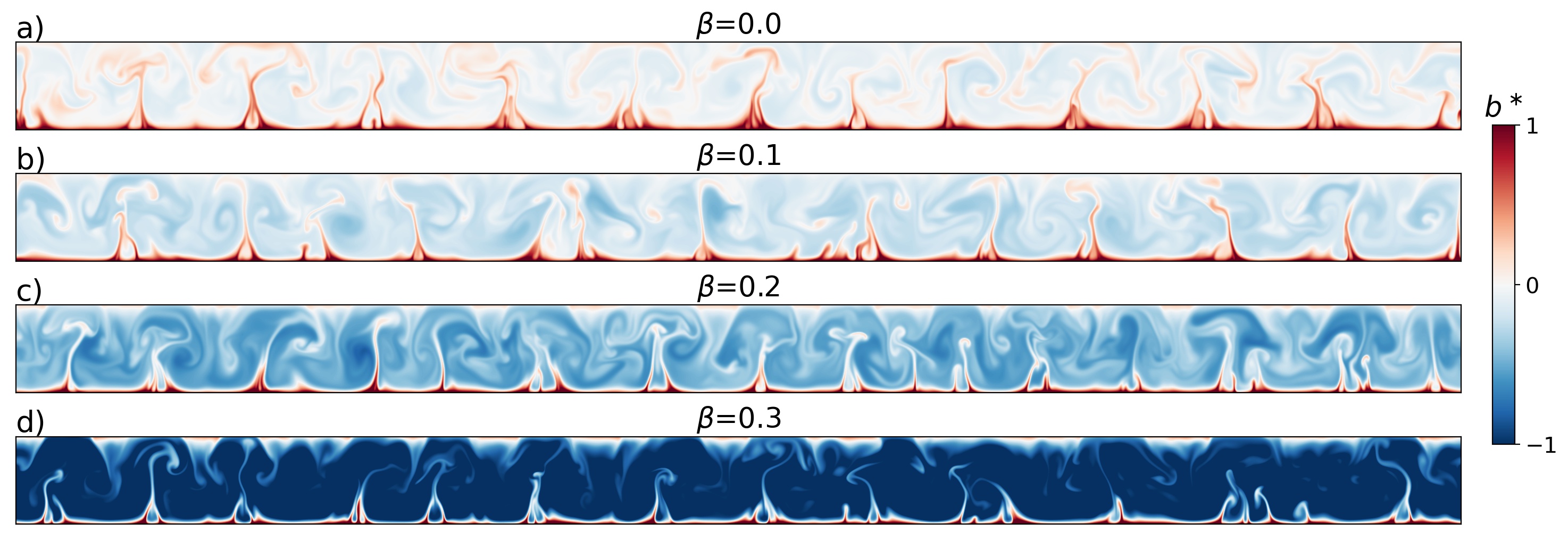}
    \caption{Instantaneous snapshot of the normalized buoyancy field $b^\ast =(b-\LTmean{b}(z=1))/\Tmean{\Delta b}$ in the statistically stationary state. The four different top boundary conditions ($\beta = 0.0,0.1,0.2,0.3$) differ in their width of the top thermal boundary layer. For the adiabatic top $\beta = 0.0$ no such layer is present.  Note that with increasing $\beta$ the range of $b^\ast$ increases.}
    \label{fig:buoyancy_field}
\end{figure*}

\begin{figure*}
\centering
    \includegraphics[width = 1.\textwidth]{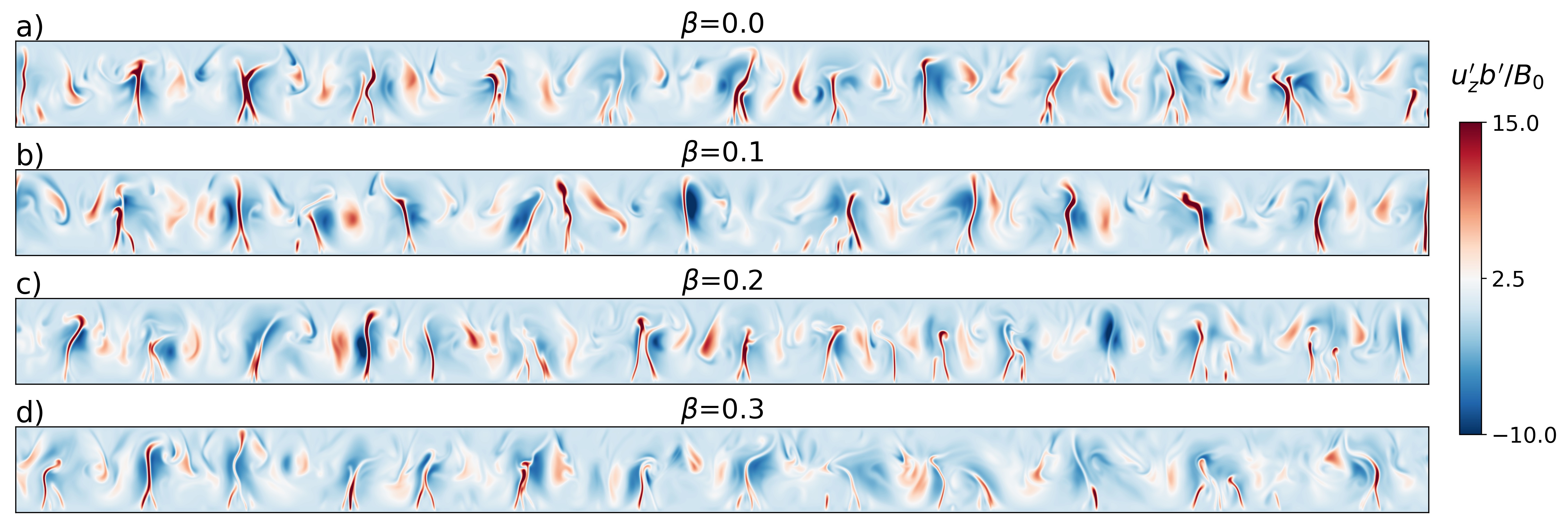}
    \caption{Instantaneous snapshot of the vertical buoyancy flux $u_z^{\prime}(x,z,t_0)b^{\prime}(x,z,t_0)$ in the statistically stationary state. The cellular order is increasingly dissolved with growing parameter $\beta$.}
    \label{fig:buoyancyflux_field}
\end{figure*}

\begin{figure}[!htbp]
    \centering
    \includegraphics[width = .35\textwidth]{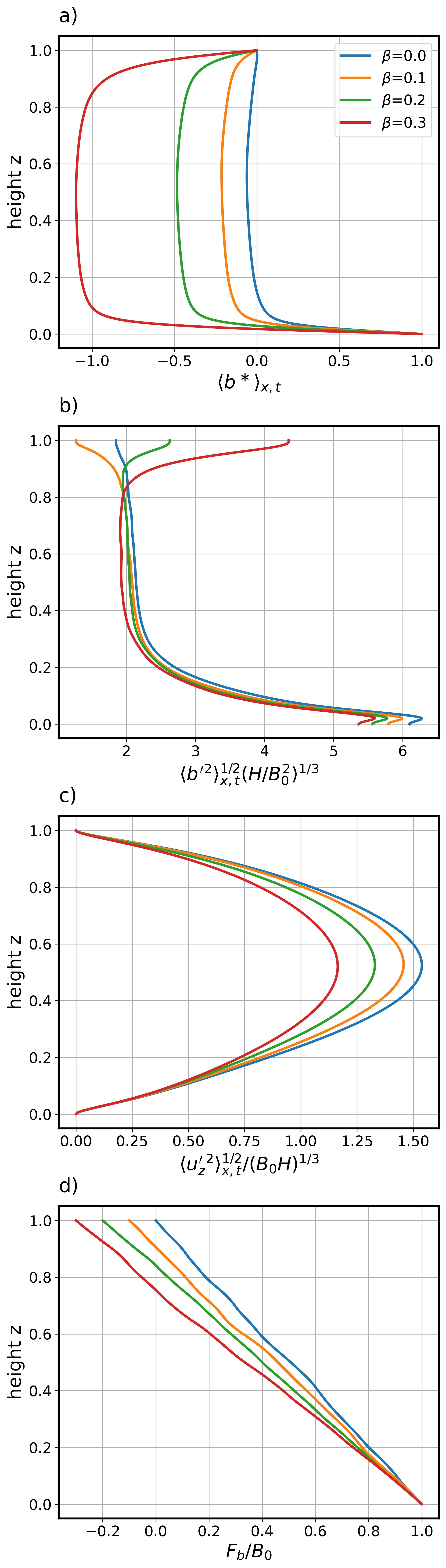}
    \caption{Vertical profiles of (a) the normalized buoyancy $b^\ast =(b-\LTmean{b}(z=1))/\Tmean{\Delta b}$ , (b) buoyancy fluctuations, (c) vertical velocity fluctuations, (d) normalized total buoyancy flux $F_b(z)/F_b(z=0)=F_b(z)/B_0$. While the boundary conditions significantly affect the buoyancy and its fluctuations, the influence on the vertical velocity profiles is less important. The fluxes show linear variation across the cell. The legend shown in a) is valid for all graphs shown.}
    \label{fig:vertical_profiles}
\end{figure}

Figures \ref{fig:buoyancy_field} and \ref{fig:buoyancyflux_field} show snapshots of the normalized buoyancy, which is given by 
\begin{equation}
 b^\ast(x,z) =\frac{b(x,z)-\LTmean{b}(z=1)}{\Tmean{\Delta b}}\,,   
\end{equation}
and the vertical flux $u_z^\prime(x,z)b^\prime(x,z)$ in the statistically stationary regime. For $\beta = 0.0$ the flux at the top is zero and no thermal boundary layer is present. This changes when the warming flux at the top becomes greater than zero, i.e. $\beta > 0$. With increasing warming flux at the top we find a thermal boundary layer at $z=1$, which increases in thickness as $\beta$ increases. Naturally, the structures in the buoyancy flux are also affected by the change of the top flux. As more buoyant fluid is transported from the top into the center of the turbulent region, the cellular order is increasingly dissolved which can be seen by prominent thermal plumes in both figures; compare panels (a) and (d). 
%------------------------------------------------------
%------------------------------------------------------
We show the line-time average vertical profiles $\LTmean{\cdot}(z)$ of $b^\ast$ in figure \ref{fig:vertical_profiles}(a). All profiles show the tendency towards a constant mean value in the central part of the domain, implying a layer of well-mixed fluid. Contrary to the common Rayleigh-B\'enard case with constant-buoyancy boundary conditions~\citep{Chilla2012}, constant-flux boundary conditions break the top-down symmetry of the mean buoyancy profile. Furthermore, for $\beta>0$, the incoming warming flux at $z=1$ results in positive buoyancy gradients and hence a stable layer at the top. 
%------------------------------------------------------
%------------------------------------------------------
We examine the variability of the velocity and buoyancy fields fields and decompose both into their volume mean $\Vmean{u_x}$, $\Vmean{u_z}$, $\Vmean{b}$ and their fluctuations $u_x', u_z', b'$
%---------------------------
\begin{align}
    \label{eq:ux_decomp}
    u_x(x,z,t) &= \Vmean{u_z}(t) + u_z'(x,z,t)\,,\\
    \label{eq:uy_decomp}
    u_z(x,z,t) &= \Vmean{u_z}(t) + u_z'(x,z,t)\,,\\
    \label{eq:b_decomp}
    b(x,z,t) &= \Vmean{b}(t) + b'(x,z,t)\,.
\end{align}
%---------------------------
Note that $\Vmean{b}$ depends on time, as it incorporates the linear warming of the fluid. Meanwhile, $\Vmean{u_x}$ and $\Vmean{u_z}$ are statistically stationary and vary weakly about their zero mean. The vertical profiles of the root mean square (r.m.s.) of $u_z'$ and $b'$ are shown in figures \ref{fig:vertical_profiles}(b) and (c). The r.m.s. of the fluctuations of the buoyancy differ greatly in their magnitude and trend in the upper portion of the domain. The vertical r.m.s. velocity component $\LTmean{u_z'^2}^{1/2}$, on the other hand, does not vary too much while changing $\beta$. Additionally, the total buoyancy flux 
\begin{align}
    F_b = \LTmean{u_z' b'} - \kappa \frac{\partial \langle b\rangle_{x,t}}{\partial z}
\end{align} 
normalized by its bottom value is shown in figure \ref{fig:vertical_profiles}(d). We find that the flux decreases linearly with increasing height. As indicated by figure \ref{fig:vertical_profiles}(a), the molecular terms mostly contribute to the near-wall regions. The turbulent transport (not shown), on the other hand, declines linearly over the middle of the domain and results in negative contributions near the top. This is expected by the stabilization by entrainment warming in the CBL~\citep{Stull:1988,Wyngaard:2010},  here considered by imposing the negative buoyancy flux $B_1$ at the top boundary. One goal of this study is to ascertain the capability to reproduce these vertical profiles of the turbulent contributions by the recurrent neural network which will be presented in the next section. 

In the following, we use the DNS data of $\beta = 0.1$ to train a recurrent neural network and make subsequent predictions for unseen data with flux ratios $\beta = 0.2,0.3$, respectively. This is done to explore the generalization properties of the echo state networks. We therefore interpolate all fields from the non-uniform grid with $2400\times 150$ points to a $720\times 30$ uniform grid by cubic splines. This grid will be denoted as the coarse-grained grid, the data as coarse-grained DNS data. 

%-----------------------------------------------------------------------------------------------------------

%---------------------------------------------------------------------
%                     Machine Learning 
%---------------------------------------------------------------------
\section{Convection prediction from echo state network}
{\label{sec:machine_learning}}
\subsection{Echo state network and echo state property}
\label{subsec:esn}

%------------------------------------------------------
%------------------------------------------------------
\begin{figure}
    \centering
    \includegraphics[width =.5\textwidth]{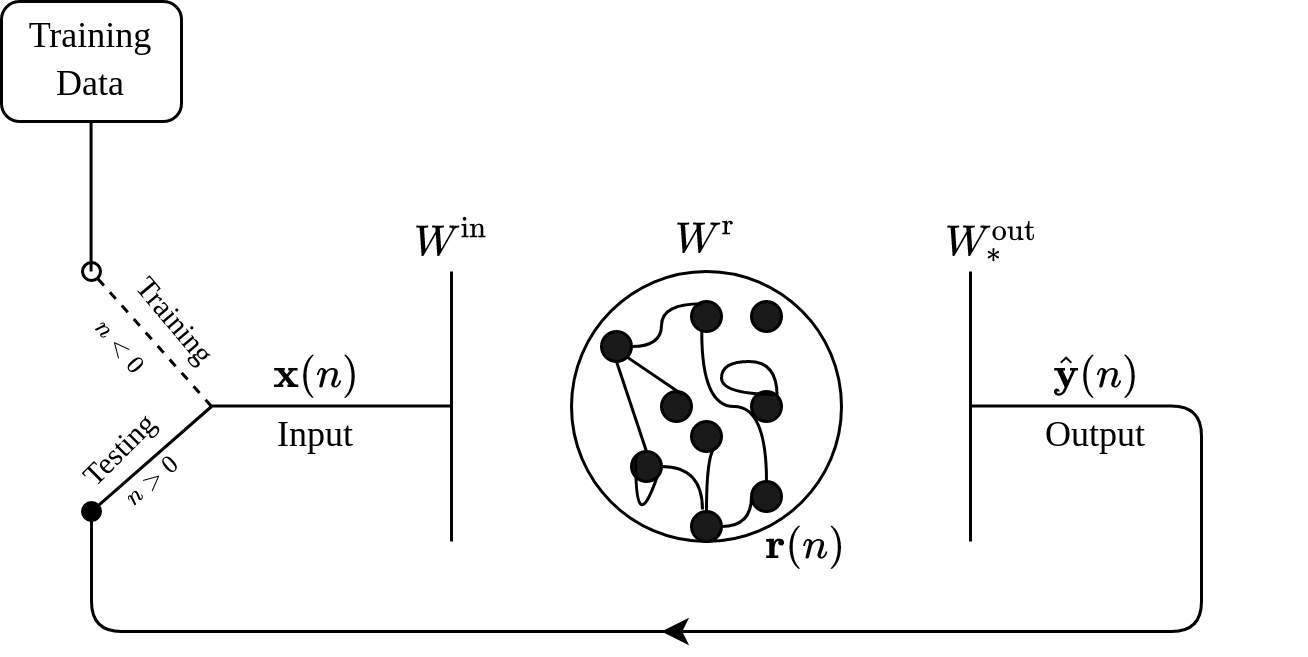}
    \caption{Echo state network architecture. For time $n<0$ the network learns the dynamics of the training data, by computing the output weights in $W_\ast^{\rm out}$. In the testing phase ($n>0$) it runs in a mode of autonomous prediction, where the last network output is fed back to the input layer to be used as new input.} 
    \label{fig:rcm_concept}
\end{figure}
%------------------------------------------------------
%------------------------------------------------------
In the following, we specify the architecture of the ESN that will be applied to process the DNS data of the CBL model described in the previous section. The reservoir state dynamics is given by
%------------------------------------------------------
\begin{align}
	\label{eq:rcm_activation}
	\mathbf{r}(n) =& (1-\gamma)\mathbf{r}(n-1) +\nonumber\\
	              & \gamma\tanh\left[W^{\rm r} \mathbf{r}(n-1) + W^{ \rm in}\mathbf{x}(n) +  d\textbf{1}\right].
\end{align}
%------------------------------------------------------
where $\mathbf{r}(n)\in \mathbb{R}^{N_{\rm r}}, \mathbf{x}(n)\in \mathbb{R}^{N_{\rm in}}$ are the reservoir state and input at time step $n$ respectively. $W^{\rm r} \in \mathbb{R}^{N_{\rm r}\times N_{\rm r}}, W^{\rm in}\in \mathbb{R}^{N_{\rm in}\times N_{\rm r}}$ are the reservoir and input weight matrices and $\gamma\in \left[0,1\right]$, $d$ are the constant leaking rate and constant bias. The reservoir output $\hat{\mathbf{y}}\in \mathbb{R}^{N_{\rm in}}$ is computed by a linear mapping of the extended reservoir state $\tilde{\mathbf{r}}(n) = \left[d,\mathbf{x}(n), \mathbf{r}(n)\right]$ (vertical concatenation of bias, reservoir input and state)
%------------------------------------------------------
\begin{equation}
	\label{eq:rcm_output}
	\hat{\mathbf{y}}(n)= W_\ast^{\rm out} \tilde{\mathbf{r}}(n).
\end{equation}
%------------------------------------------------------
The fitted output weights $W_\ast^{\rm out}\in \mathbb{R}^{N_{\rm in}\times (1+N_{\rm in}+N_{\rm r})}$ are chosen as to minimize the mean square cost function
%------------------------------------------------------
\begin{equation}
	\label{eq:rcm_costfunction}
	C(W^{\rm out}) = \sum\limits_{n=-T_L}^{-1}\| \mathbf{y}(n) - W^{\rm out}\tilde{\mathbf{r}}(n)\|_2^2 + \lambda\|w_i^{\rm out}\|_2^2,
\end{equation}
%------------------------------------------------------
where $\mathbf{y}$ are the target outputs, which are part of the training data. $T_L$ is the number of training time steps, $w_i^{\rm out}$ is the $i^{\rm th}$ row of $W^{\rm out}$ and $\|\cdot\|_2$ denotes the $L^2$ norm. The last term penalizes large values of the rows of the output weight matrix by adjusting the regression parameter $\lambda$. This concept is one possibility to counter the problem of overfitting, where the machine learning algorithm learns the training data by heart, consequently performing poorly when operating on data outside the training data set. The solution to this $L^2$-penalized linear regression problem is given by
%------------------------------------------------------
\begin{equation}
	\label{eq:rcm_linearreg}
    W_\ast^{\rm out} =Y R^T \left(RR^T+\lambda I\right)^{-1}
\end{equation}
%------------------------------------------------------
where the $n^{\rm th}$ column of $Y\in \mathbb{R}^{N_{\rm in}\times T_L}$, $S\in \mathbb{R}^{N_{\rm r}\times T_L}$ are $\mathbf{y}(n)$ and $\tilde{\mathbf{r}}(n)$ respectively. $I\in \mathbb{R}^{N_{\rm r}\times N_{\rm r}}$ denotes the identity matrix and $(\cdot)^T$, $(\cdot)^{-1}$ are the transpose and inverse. After the training phase an initial input is given at $n = 0$ and the reservoir output at time step $n\ge 0$ is fed back to the input layer, by letting $\mathbf{x}(n) = W_\ast^{\rm out} \mathbf{r}(n-1)$. During this testing phase the ESN autonomously predicts the next $T_T$ iterations of the initial input. Figure \ref{fig:rcm_concept} summarizes the architecture of the ESN in a sketch.

This inexpensive training procedure comes at a cost of finding a suitable set of hyperparameters, i.e. parameters which are not learned and have to be tuned beforehand. Here we restrict ourselves to ${\rm h} = \{\gamma, \lambda, N_{\rm r}, D, \varrho\}$. The last two quantities are the reservoir density $D$ and spectral radius $\varrho$. They are algebraic properties of the reservoir weight matrix and represent the number of non-zero elements and largest absolute eigenvalue of $W^{\rm r}$, respectively. Finding a right setting of these hyperparameters is crucial, as they influence the memory capacity of the reservoir~\citep{Hermans2010}. In \cite{jaeger__2001} a necessary condition for an effective reservoir was proposed: the \textit{echo state property}. A reservoir is said to possess echo states when two different reservoir states $\mathbf{r}_1(n-1)$, $\mathbf{r}_2(n-1)$ converge to the same reservoir state $\mathbf{r}(n)$, provided the same input $\mathbf{x}(n)$ is given and the system has been running for many iterations $n$. This property highly depends on the data one uses, a suitable set of hyperparameters ${\rm h}$, as well as the reservoir initialization~\citep{lukosevicius_practical_2012}. So far, no universal rule for the presence of echo states has been proposed. On top of that, the echo state property is merely a necessary condition and no feasible sufficient condition has yet been found as discussed in \cite{Yildiz2012}. We will keep using reservoir initializations and hyperparameter ranges, which have shown good results, e.g., in \cite{Pandey2020} or \cite{Heyder2021}. We initialize the input and reservoir weights as random, i.e., $W^{\rm in}\sim \mathcal{U}\left[ -0.5, 0.5\right]$ and $W^{\rm r}\sim \mathcal{U}\left[ 0, 1\right]$. $W^{\rm r}$ is then normalized by its largest absolute eigenvalue and is subsequently scaled by $\varrho$. Afterwards, randomly selected entries of this matrix are set to zero to assure the specified value of the reservoir density $D$ is obtained. The specific value of each of the quantities in $h$ is chosen by a grid search procedure which will be discussed further below.

\subsection{Network training with case at $\beta=0.1$}
\label{subsec:network_training}
%------------------------------------------------------
\begin{figure*}[!htpb]
    \centering
    \includegraphics[width = .7\textwidth]{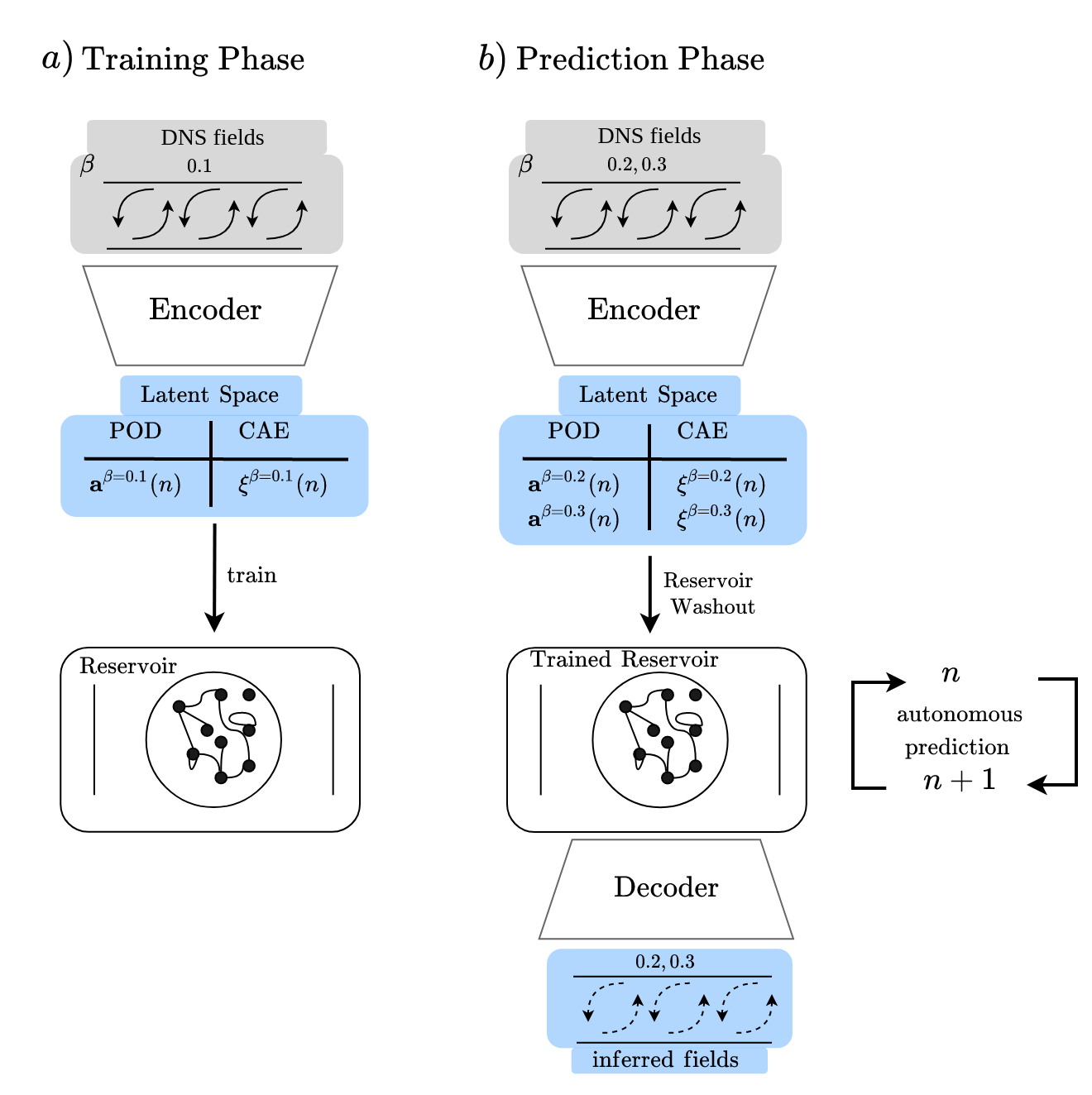}
    \caption{Sketch of the transfer learning concept. (a) During the training phase, 700 snapshots of the simulation data for $\beta=0.1$ are encoded into the latent space, either via the reduction by POD (denoted as $\textbf{a}^{\beta=0.1}$) or via the one by a CAE (denoted as $\xi^{\beta=0.1}$). A reservoir is subsequently trained with the latent space. The network learns the dynamics; the optimal output weights are obtained. (b) In the prediction phase, the reservoir is then used to infer the dynamics of the target latent spaces at $\beta = 0.2, 0.3$ and predicts either $\textbf{a}^{\beta=0.2,0.3}$(POD) or $\xi^{\beta=0.2,0.3}$ (CAE). Snapshots of the convection flow can then be reconstructed and validated by the corresponding decoder to obtain fully resolved fields for the cases of $\beta = 0.2$ and $\beta = 0.3$.}
    \label{fig:rcm_transfer_concept}
\end{figure*}
%------------------------------------------------------
In the following, we explore whether we can use the ESN to infer changes in the convective flow, induced by changes in the buoyancy flux at the top of the two-dimensional domain. A trained network is thus exposed to unseen data at a different physical parameter set. Such a procedure probes the generalization properties of the ESN. The subject is also connected to a transfer of the learned parameters from one task to a similar one which is known as transfer learning~\citep{Pan2010}. Due to the computationally inexpensive training scheme of ESNs, transfer learning is not often applied for this class of algorithms, even though implementations have been proposed very recently~\citep{Inubushi2020}. 

Here, we take a different approach which is sketched in figure \ref{fig:rcm_transfer_concept}. A reservoir is trained with the reduced data of one case of buoyancy boundary conditions at $z=1$, namely $\beta=0.1$. Finally, we use the trained network for predicting the dynamics and statistical properties of two different and unseen convective flows with buoyancy flux parameter $\beta = 0.2$ and $0.3$.

The DNS data possesses many degrees of freedom, so that we have to introduce a preprocessing step before passing the convection data to the reservoir. We propose two common reduced order modelling techniques, the (1) Proper Orthogonal Decomposition (POD) and the (2) Convolutional Autoencoder (CAE). The former is well known in fluid mechanics as a linear method, where the data reduction is realized by a truncation to a set of Galerkin modes \cite{Sirovich1987}. The CAE on the other hand, represents a deep convolutional neural network, commonly used in deep learning tasks, such as feature extraction in image processing \cite{Baldi2012}. For brevity we only mention major aspects of both methods here and move details to the appendix.

For both data reduction approaches we sample $700$ time steps of our coarse-grained DNS data in an interval of $0.25T_f$ for the simulation of $\beta = 0.1$ in the statistically stationary regime. Also, snapshots of $700$ further time steps with the same sampling interval are gathered for the unseen target simulations at $\beta = 0.2$ and $0.3$. Before reducing the dimensionality of the data, we decompose the buoyancy fluctuations further
%------------------------------------------------------
\begin{equation}
    \label{eq:b_decomp2}
    b'(x,z,t) = \Tmean{b'}(x,z)+ b''(x,z,t).
\end{equation}
%------------------------------------------------------
Finally we apply both POD and CAE on the vector $\mathbf{g} =(u_x', u_z', b'')^T$. Both methods are chosen to reduce the dimensionality of this vector to $N_{\rm POD} = N_{\rm CAE} = 300$ features per snapshot. The total number of degrees of freedom is thus reduced from three fields on a grid with size $2400\times 150$ in the original DNS (that corresponds to $N_{\rm dof}=1.08\times 10^6$) via coarse grained data of grid size $720\times 30$ to $300$ modes in the latent space by a factor of $3600$. With this choice of $N_{\rm POD}$ the POD reduction captures about $80\%$ of the original energy (for more details see appendix). 

We refer to this reduced data as POD time coefficients $\mathbf{a}(n) = (a_1(n), a_2(n)$, $...$, $a_{N_{\rm POD}}(n))^T$ for the data reduction via POD and as encoding space $\mathbf{\xi}(n) = (\xi_1(n), \xi_2(n)$, $...$, $\xi_{N_{\rm CAE}}(n))^T$ for the one via CAE.

We construct the training data set for our ESN by taking $700$ instances of $\mathbf{a}(n)$ or $\mathbf{\xi}(n)$ of $\beta = 0.1$. This results to a total training length of $T_L = 700$. During this phase the reservoir is trained to predict the respective next time instance of the POD expansion coefficients $\mathbf{a}(n+1)$ or encoding variables $\mathbf{\xi}(n+1)$, see again eq.(\ref{eq:rcm_linearreg}).

In the next paragraph, we explain how these trained ESNs can be used to predict the time coefficients (or encoding space) of the two cases with different heat flux parameter, namely $\beta = 0.2,0.3$. Finally in \ref{subsec:inferring_beta_esn} the individual prediction performance of both POD and CAE method together with the ESN will be examined.

\subsection{Prediction for unseen cases at $\beta=0.2$ and $0.3$}
\label{subsec:inferring_beta_esn}
Once the ESN has learned to process the data in the latent space (which are obtained either by POD or CAE) for the case of $\beta=0.1$, it is exposed to unseen data of the two CBL model cases, $\beta = 0.2, 0.3$ without further training adjustments. For this, we initialize a new reservoir state which is preceded by 50 iterations of Eq.(\ref{eq:rcm_activation}), where the reservoir input is given by 50 time steps of either $\mathbf{a}^{\beta = 0.2}$ and $\mathbf{a}^{\beta=0.3}$ in case of reduction by POD or $\xi^{\beta = 0.2}$ and $\xi^{\beta=0.3}$ in case of CAE, see also figure \ref{fig:rcm_transfer_concept}(b). With this washout phase, we intend to transition to the new parameter regime of $\beta = 0.2$ or $0.3$. Starting from this reservoir state, the ESN will autonomously predict $T_T = 700$ future time steps with its output weights that were learned for $\beta=0.1$. We validate these predictions by a direct comparison with $\mathbf{a}^{\beta=0.2}(n)$, $\mathbf{a}^{\beta=0.3}(n)$ and $\xi^{\beta=0.2}(n)$, $\xi^{\beta=0.3}(n)$ with $n\in \left[1,T_T\right]$, respectively. For this we apply the {\em mean squared prediction error} (MSE) which, e.g., for the specific case of $\beta = 0.2$ is given by
\begin{equation}
    \label{eq:mse}
    {\rm MSE}_{\rm h} = \frac{1}{T_T}\sum\limits_{n=1}^{T_T}\| \hat{\mathbf{y}}(n)- \mathbf{a}^{\beta=0.2}(n)\|_2^2\,.
\end{equation}
In addition, we take the {\em normalized average relative error} (NARE) of the reconstructed fields $u_z'$, $b''$ and $u_z'b''$. The definition follows the work of~\cite{srinivasan2019} and is given for example for $u_z'b''$ by 
%------------------------------------------------------
\begin{align}
	E_{\rm h}\left[\langle u_z'b''\rangle_{x,t}\right] &= \frac{1}{C_{\max}}\int\limits_0^1 \Big|\langle u_z'b''\rangle_{x,t}^{\rm ESN}(y) -\langle u_z'b''\rangle_{x,t}^{\rm POD}(z)\Big| dz
	\label{eq:nare}
	\intertext{with}
	C_{\max}&=\frac{1}{2B_0 \max_{z\in \left[0,1\right]}(\left|\langle u_z'b''\rangle_{x,t}^{\rm POD}\right|)}.
\end{align}
%------------------------------------------------------
The superscript indicates whether the field is reconstructed, see Eq. (\ref{eq:pod_rom}), from the $N_{\rm POD}$ POD time coefficients (POD) or the ESN predictions (ESN). This measure quantifies errors in the line-time average profiles of the physical fields. Similarly, one can define MSE and NARE for the CAE case by using $\mathbf{\xi}$ instead of $\mathbf{a}$ and the CAE instead of the POD reconstruction.

Our choice of the optimal ESN hyperparameters ${\rm h}_\ast$ is listed in table \ref{tab:hyperparam}. We conducted grid searches of $N_{\rm}$, $D$, $\gamma$ and $\varrho$. See the appendix for more details. For each setting, we additionally took $100$ random realizations of the same reservoir setting and computed ${\rm MSE}_{\rm h}$ and $E[u_z'b'']$. The final setting ${\rm h}_\ast$ was chosen according to the lowest third quartile of $E[u_z'b'']$ of all 100 samples. We deliberately choose the third quartile over the median, as it assures robust reservoir outputs for different random weights $W^{\rm in}$, $W^{\rm r}$ and therefore more reliable predictions. Furthermore, we choose the NARE of the buoyancy flux, due to its physical relevance, as opposed to the MSE. Moreover, it is comprised of two quantities which are prone to prediction errors.
%------------------------------------------------------
\begin{table}[!htpb]
    \renewcommand{\arraystretch}{1.5} 
    \centering
    \begin{tabular}{ccccccc}
        \hline\hline
         & $\beta$ & $\gamma$ & $\lambda$ & $N_{\rm r}$ & $D$ & $\varrho$ \\
        \hline
        POD$\quad$ & $0.2$ & $0.8$ & $0.5$ & $1024$ & $0.84$ & $1.60$ \\
        POD$\quad$ & $0.3$ & $0.8$ & $0.5$ & $1024$ & $0.84$ & $1.85$ \\
        CAE$\quad$ & $0.2$ & $0.2$ & $0.5$ & $1024$ & $0.84$ & $1.98$ \\
        CAE$\quad$ & $0.3$ & $0.2$ & $0.5$ & $1024$ & $0.84$ & $0.81$ \\
        \hline\hline  
    \end{tabular}
    \caption{Choice of optimal ESN hyperparameters ${\rm h}_\ast$. The values were chosen according to a grid search. See the appendix C for more detailed information on the grid search.}
    \label{tab:hyperparam}
\end{table} 
%------------------------------------------------------
\subsubsection{Results for Proper Orthogonal Decomposition-Echo State Network}
\begin{figure*}[!htbp]
    \centering
    \includegraphics[width = \textwidth]{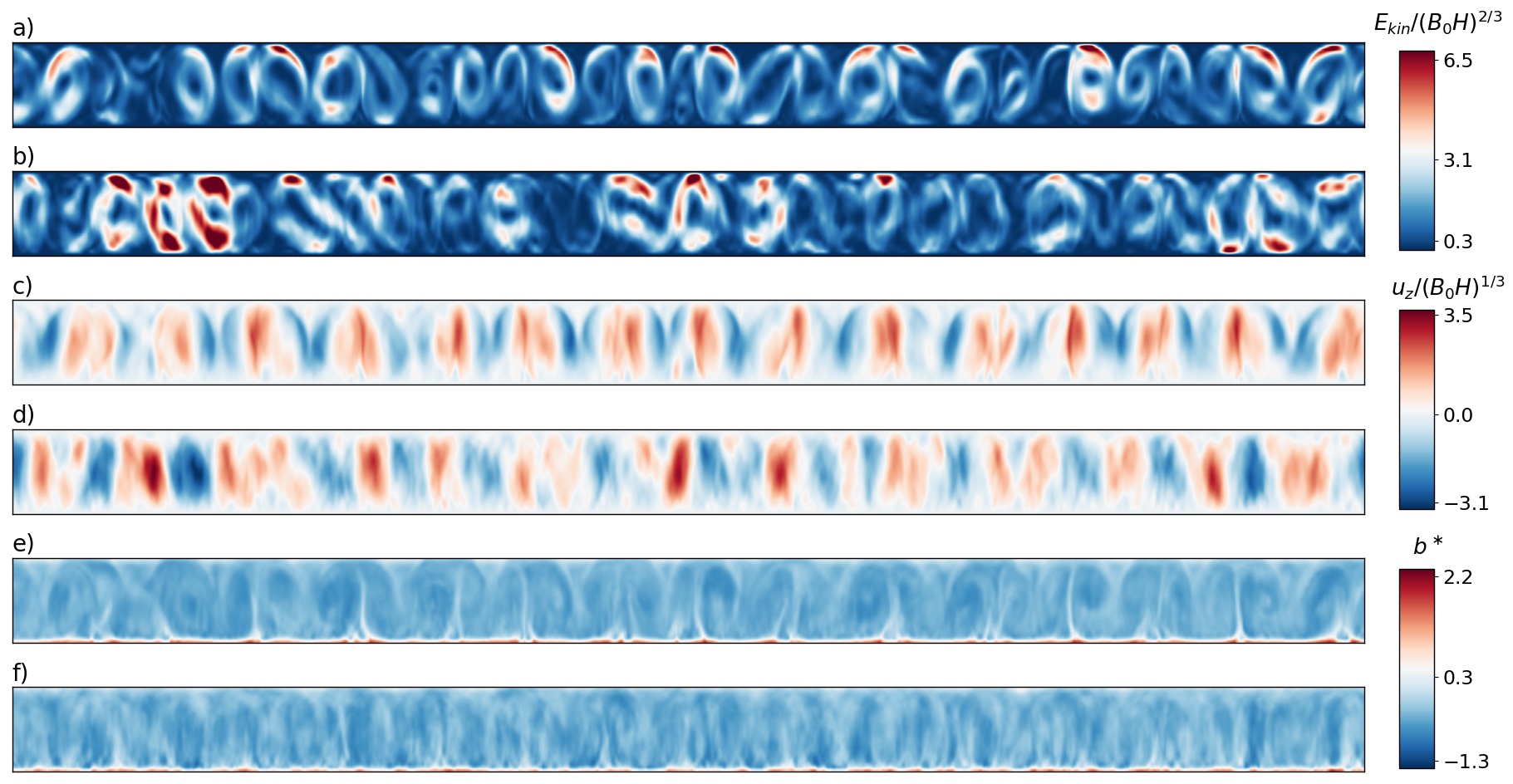}
    \caption{POD case for inferring $\beta = 0.2$. Instantaneous snapshots of the local turbulent kinetic energy $E_{\rm kin}(x,z,t_0)$ in panels (a,b), the vertical velocity component $u_z$ in panels (c,d) and the normalized buoyancy $b^\ast$ in panels (e,f) at time step $n=350$ in the prediction phase. POD reconstructions with the most energetic $N_{\rm POD}$ modes of $\beta = 0.2$ (validation snapshot) are shown in panels (a), (c), and (e). The corresponding ESN predictions are displayed in panels (b), (d), and (f).}
    \label{fig:results_fields_pod}
\end{figure*}

We reconstruct each component of the physical fields $u_x, u_z$ and $b$ via eq. \eqref{eq:pod_rom} using the decompositions  \eqref{eq:ux_decomp} -- \eqref{eq:b_decomp} and (\ref{eq:b_decomp2}). For the validation, we use the expansion coefficients of the first $N_{\rm POD}$ modes of the $\beta = 0.2$ and $\beta = 0.3$ data. For $\beta=0.2$, instantaneous snapshots in the middle of the prediction phase (the time step is $n=350$) of the {\em local} turbulent kinetic energy 
\begin{equation}
E_{\rm kin}(x,z,t) =\frac{1}{2}[u_x^2(x,z,t)+u_z^2(x,z,t)]\,,
\end{equation} 
the vertical velocity component $u_z(x,z)$, and the normalized buoyancy $b^\ast(x,z)$ can be seen in figure \ref{fig:results_fields_pod}. The ground truth, i.e. the POD data, is shown for comparison. We find common features in the predicted and the validation fields. Even though some magnitudes deviate, roll patterns in the kinetic energy can be identified in the prediction case. In the velocity field component, vertical up- and downdrafts can be clearly identified. Their width and shape differs slightly from the ground truth. Moreover, the thermal boundary layer at $z=1$ is reproduced in the predicted buoyancy field. Thermal plumes which detach primarily from the bottom wall can also be identified. It is clear that some features are not perfectly reproduced, but the qualitative picture agrees fairly well.
%------------------------------------------------------
%------------------------------------------------------
We emphasize that these results were obtained for one particular realization out of the 100 reservoirs with the same hyperparameter setting, that were taken typically. Nevertheless, both results are exemplary for their setting ${\rm h}_\ast$, as they correspond to the median NARE of the buoyancy flux. 

\begin{figure}[!htpb]
    \centering
    \includegraphics[width = .35\textwidth]{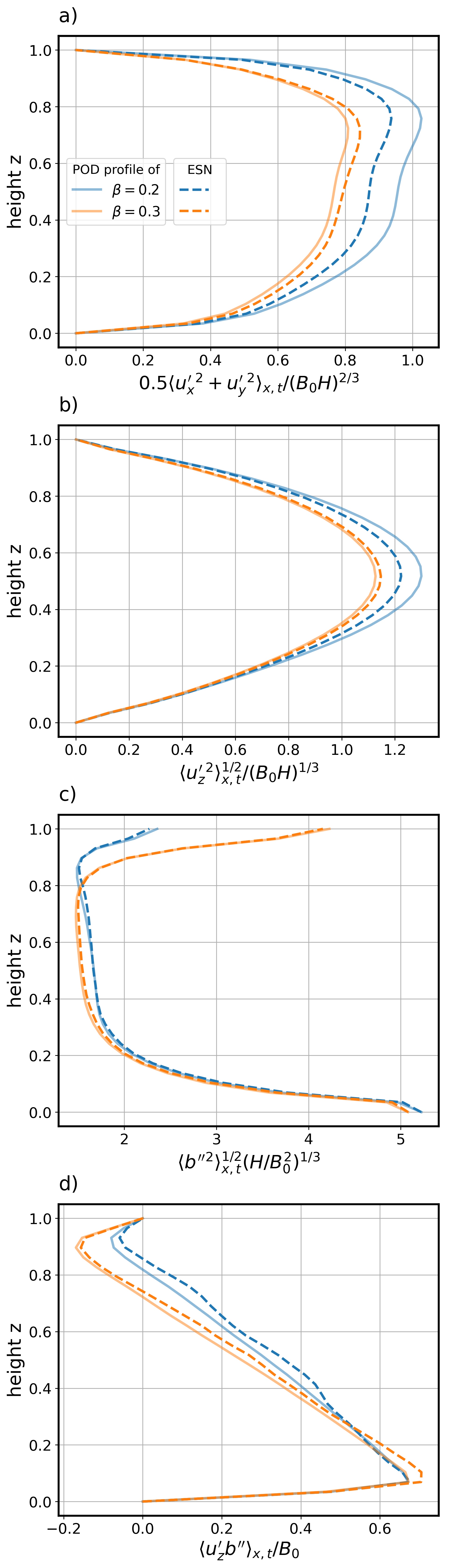}
    \caption{Line-time average profiles. (a) Turbulent kinetic energy. (b) Root mean square profile of vertical velocity fluctuations. (c) Root mean square buoyancy fluctuations profile of  $b''$. (d) Convective buoyancy flux. The ESN predictions (dotted lines) were chosen as to hold the median buoyancy flux NARE. They reproduce some low-order statistics of the truncated POD reconstruction (solid lines) of $\beta = 0.2$ (blue) and $\beta = 0.3$ (orange).}
    \label{fig:results_LTA_pod}
\end{figure}

We now investigate the generalization capability of the reservoir by computing line-time average profiles $\LTmean{\cdot}$ of the fluctuations of the corresponding fields for $\beta=0.2$ and $0.3$. These are important parameters of simulations of large-scale turbulence. The profiles are given in figure \ref{fig:results_LTA_pod} and their corresponding NARE values are listed in table \ref{tab:results_nare}. We find that in this setting ${\rm h}_\ast$ the average reservoir produces reasonable approximations to the profiles of the true low-order statistics of the both $\beta$ values. Despite some deficiency in the profiles of turbulent kinetic energy and vertical velocity for $\beta = 0.3$, the asymmetry due to the boundary conditions is captured in all profiles. Especially the buoyancy fluctuations are reproduced well. While the ESN reproduces the linear decrease of the convective buoyancy flux, $\LTmean{u_z'b''}^{\rm ESN}$, it overshoots near the bottom of the cell for $\beta = 0.2$ as well as in the upper cell for $\beta = 0.3$ . Nevertheless, the inferred profiles match the ground truth to a reasonable extent.
%------------------------------------------------------
%------------------------------------------------------
\begin{table*}[!hptb]
    \centering
    \begin{tabular}{cccccc}
         \hline\hline
          & $\beta$ & $E_{h_\ast}\left[0.5\LTmean{u_x'^2+u_z'^2}\right]$ & $E_{h_\ast}\left[\LTmean{u_z'^2}\right]$ & $E_{h_\ast}\left[\LTmean{b''^2}\right]$ & $E_{h_\ast}\left[\LTmean{u_z'b''}\right]$ \\
         \hline
         POD$\quad$ & $0.2$ & $0.033$ & $0.024$ & $0.002$ & $0.018$\\
         POD$\quad$ & $0.3$ & $0.015$ & $0.009$ & $0.003$ & $0.018$\\
         CAE$\quad$ & $0.2$ & $0.053$ & $0.040$ & $0.007$ & $0.013$\\
         CAE$\quad$ & $0.3$ & $0.027$ & $0.029$ & $0.017$ & $0.035$\\
         \hline\hline
    \end{tabular}
    \caption{Normalized average relative errors, see eq.~(\ref{eq:nare}), of the inferred line-time average profiles shown in figures \ref{fig:results_LTA_pod} and \ref{fig:results_LTA_cae} for the reduction by POD and CAE, respectively.}
    \label{tab:results_nare}
\end{table*}
%------------------------------------------------------
Overall, the ESN generalizes well to unseen convection data with similar boundary conditions, when using the low-order POD model. The inferred fields of $\beta=0.2$ and $0.3$ (see appendix) reproduce features like thermal boundary layer, up- and downdrafts as well as roll patterns. In the next section, we investigate how the ESN performs when we combine it with a trained convolutional autoencoder.

\subsubsection{Results for Convolutional Autoencoder-Echo State Network}
\begin{figure*}[!htpb]
    \centering
    \includegraphics[width = \textwidth]{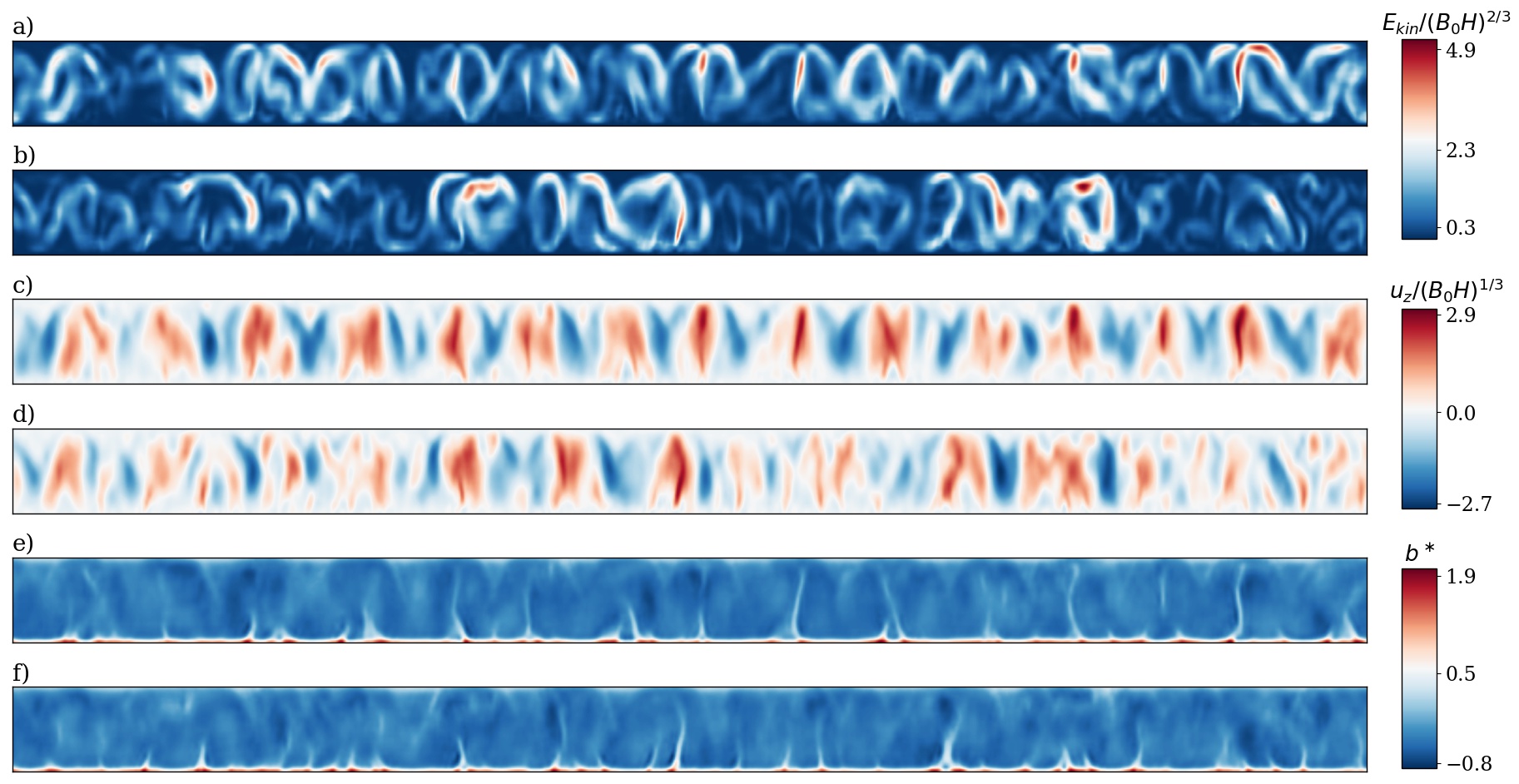}
    \caption{CAE case for inferring $\beta = 0.2$. Instantaneous snapshots of the turbulent kinetic energy $E_{\rm kin}(x,z,t_0)$ in panels (a,b), the vertical velocity component $u_z$ in panels (c,d) and the normalized buoyancy $b^\ast$ in panels (e,f) at time step $n=350$ in the prediction phase. CAE reconstructions of $\beta = 0.2$ (validation snapshot) are shown in panels (a), (c), and (e). The corresponding ESN predictions are displayed in panels (b), (d), and (f).}
    \label{fig:results_fields_cae}
\end{figure*}

\begin{figure}[!htpb]
    \centering
    \includegraphics[width = .35\textwidth]{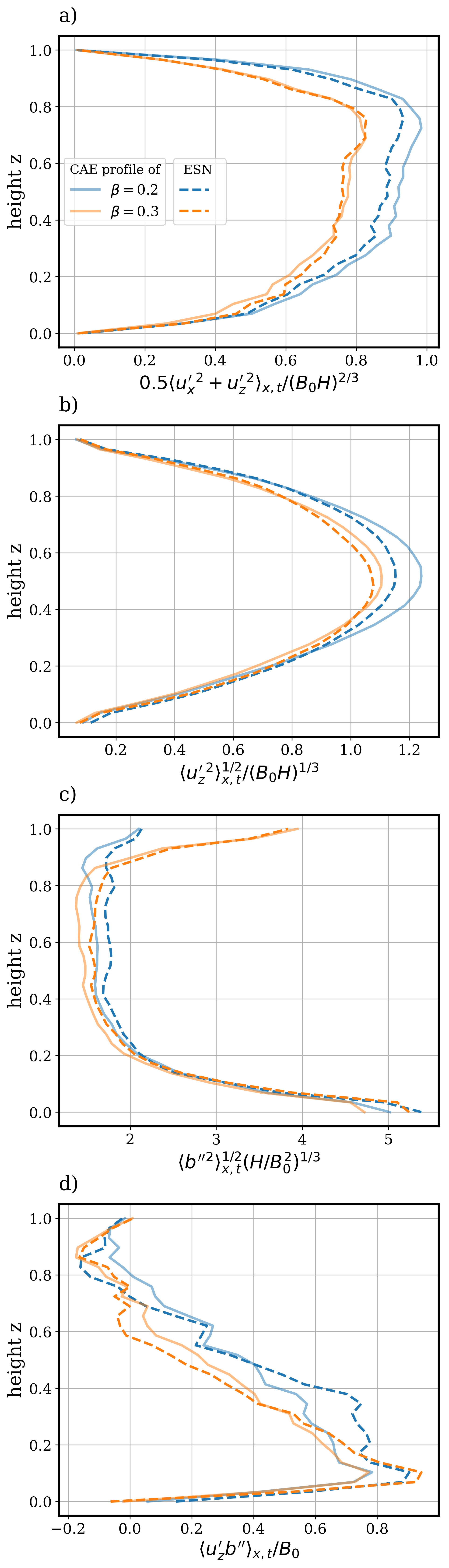}
    \caption{Line-time average profiles. (a) Turbulent kinetic energy. (b) Root mean square profile of vertical velocity fluctuations. (c) Root mean square buoyancy fluctuations profile of  $b''$. (d) Convective buoyancy flux. The autoencoder introduces artefacts in the statistical profiles (solid lines) of $\beta = 0.2$ (blue) and $\beta = 0.3$ (orange).The ESN predictions (dotted lines) were chosen as to hold the median buoyancy flux NARE.}
    \label{fig:results_LTA_cae}
\end{figure}

By decoding the inferred latent spaces using eq.~\eqref{eq:decoder} and eqns.~\eqref{eq:ux_decomp}--\eqref{eq:b_decomp} as well as eq.~\eqref{eq:b_decomp2}, we reconstruct the fields $u_x$, $u_z$ and $b$. Figure \ref{fig:results_fields_cae} shows instantaneous snapshots of turbulent kinetic energy, vertical velocity and normalized buoyancy of inferred fields (ESN) and ground truth (CAE). Here, we find predicted and true fields almost indistinguishable in terms of their features. Roll patterns, up- and downdrafts as well as thermal plumes detaching from the bottom are reproduced very naturally. While the POD method introduces some deviations in the inferred fields, the autoencoder reproduces the small-scale features of the convection patterns well.
%------------------------------------------------------
%------------------------------------------------------
%------------------------------------------------------
%------------------------------------------------------
Figure \ref{fig:results_LTA_cae} shows the inferred line-time averaged profiles of the physical fields. Their corresponding NARE values are listed in table \ref{tab:results_nare}. Differently to the  the linear POD method, the CAE is trained by a gradient descent procedure, which introduces artefacts in the statistical profiles (solid lines). The loss of information in the encoder-decoder structure thus impacts the statistical features of the reconstructed flow. As a consequence, larger magnitudes of this measure can be observed for most entries of the table. 

It is possible that this error can be reduced by introducing an additional term to the loss function of the CAE that penalizes large deviations from the mean profiles. This is not applied here. The biggest artefacts can be seen in the buoyancy flux in figure \ref{fig:results_LTA_cae}(d). Nevertheless, we find the differences acceptable, as the asymmetry and shape of the true profiles are retained. The reservoir manages to reproduce the overall trend of the line-time average profiles. While for $\beta = 0.3$, turbulent kinetic energy, vertical velocity and buoyancy fluctuations seem to be harder to match for the ESN, the buoyancy flux shows good agreement. The $\LTmean{u_z'b''}$ profile of the intermediate case $\beta = 0.2$ is poorly predicted in the lower boundary layer, where the maximum value is overestimated by the reservoir. 

%------------------------------------------------------
%------------------------------------------------------

%------------------------------------------------------
%------------------------------------------------------
%---------------------------------------------------------------------
%                     Conclusion
%---------------------------------------------------------------------
\section{Conclusions and outlook}
{\label{sec:conclusion}}
In this work, we explored the generalization property of a machine learning method applied to a more complex convection flow than standard RBC. In particular, we considered echo state network algorithms applied to two-dimensional convection with different buoyancy boundary conditions at the top and bottom. To this end, we impose buoyancy fluxes at the vertical boundaries which can be understood as entrainment from the top and surface heating from the bottom in an atmospheric convective boundary layer. The model is hence characterized by the buoyancy flux ratio $\beta$, beside Rayleigh and Prandtl numbers. An increasing value of $\beta$ quantifies a counter-heating that stabilizes the top layer and results in negative values of the mean convective buoyancy flux close to the top boundary. Thus our model resembles properties that are absent in a standard Rayleigh-B\'{e}nard setup with uniform temperatures at the top and bottom. In particular, the top-down symmetry of the boundary layers is broken; in this respect the present model is similar to a complex non-Boussinesq convection flow. It is thus an ideal testing bed for dynamic parametrizations of the buoyancy flux and its low-order moments by machine learning algorithms. On the other hand, it is still a simplification of an atmospheric layer, in particular in view to its two-dimensionality.

We conducted a series of direct numerical simulations for values of $\beta$ that vary between 0 and 0.3, a range that represents mid-day atmospheric conditions over land. An adiabatic top boundary, i.e., zero incoming and outgoing flux ($\beta=0$), is also considered for comparison. The four simulations result in flows with distinct features which are not common in Rayleigh-B\'enard convection. The mean buoyancy is constant throughout the middle of the domain, which resembles a mixed layer inside the convective cell. Further, positive buoyancy gradients at the top and a linear decline with height of the covariance of vertical velocity and buoyancy can be observed, both features which are also observed in an atmospheric boundary layer. The four simulations also display different dynamics and convection patterns, which demonstrates the impact of the incoming top flux. These differences become evident when considering the low-order statistics of the buoyancy and its vertical flux. As $\beta$ increases, so does the thickness and magnitude of the stable layer at the top of the convection cell, and the intensity of the buoyancy fluctuations. 

Regarding the machine learning method, we employ a recurrent neural network in the form of an echo state network to predict the dynamics and low-order statistics for the {\em unseen} simulation data at $\beta=0.2$ and $0.3$. The echo state network is trained with simulation data records at $\beta=0.1$. In this way, we can explore the generalization properties of the neural network, or in other words, the performance of the machine learning algorithm to unseen data with different physical parameters. 

We use two common approaches to reduce the amount of DNS data for the prediction task, (1) the proper orthogonal decomposition and the (2) convolutional autoencoder. Both methods reduce the data to $300$ degrees of freedom per snapshot. We find that the training of the echo state network with data of the low-magnitude flux  ($\beta = 0.1$) at the top yields good approximations of the dynamics of the higher-magnitude turbulent flux cases at $\beta = 0.2$ and $0.3$. This is the case for both data reduction methods. We are also able to reconstruct velocity and buoyancy fields very well. This is in line with a low-order statistics of these fields which is also properly reconstructed, for example for the vertical profiles of the buoyancy flux. 

We point out that the two low-order models differ in their compression technique and hence yield different performances, when combined with reservoir computing model in the latent space. While the POD preserves line-time average profiles, the autoencoder introduces small artefacts to the statistics. The quality of the predicted spatial features differs also among the methods, as the predicted POD time coefficients capture coarse convection features, while the convolutional autoencoder reproduces the natural convection patterns, i.e., the prominent features, very well. We can conclude that for our setup, data emerging from one case with constant flux boundary conditions can be used to infer at least statistical and spatial features of two different cases with different conditions. The echo state network can thus serve as a reduced-order and scalable dynamical model that generates the appropriate turbulence statistics without solving the underlying Navier-Stokes equation of the flow.

The present study can be considered as one step in the development of efficient reduced dynamical models of convection processes by machine learning methods. Several directions for the future research are possible from this point. First, an extension to the three-dimensional case is desirable. This requires a stronger reduction of the data which can be achieved by even deeper convolutional encoder-decoder networks in combination with spatial filtering of the direct numerical simulation data. Such a reduction could lead to a dynamical version of a recent approach by \cite{Fonda2019} which reduced the convective turbulent heat transport across a convection layer to a dynamic planar network. We also suggest to incorporate physical laws or known flow properties in the training routine of the autoencoder, as mere mimicking of the input fields produces artefacts in the statistical features of the reconstruction. This has been done in several works, e.g., in ref. \cite{Raissi2020}. Moreover, one should keep the balance between the demand of physical reality and computational expense as to keep the use of a low-order model meaningful. 

Furthermore, by definition neural networks are not designed to process data that live on a continuum of different lengths and times, a property which is immanent to turbulent flows. Architectures which can represent the multiscale nature of turbulence are required. Studies in these directions are currently underway and will be reported elsewhere.

\vspace{0.3cm}
\section*{Acknowledgments}
This work is supported by the project No. P2018-02-001 "DeepTurb -- Deep Learning in and of Turbulence" of the Carl Zeiss Foundation. Partial support for the second author was provided by grant PID2019-105162RB-I00 funded by MCIN/AEI/10.13039/501100011033. The authors gratefully acknowledge the Gauss Centre for Supercomputing e.V. (\url{www.gauss-centre.eu}) for funding this project by providing computing time through the John von Neumann Institute for Computing (NIC) on the GCS Supercomputer JUWELS at J\"ulich Supercomputing Centre (JSC).
\FloatBarrier

%----------------
% APPENDIX
%----------------
\appendix
\section{Proper Orthogonal Decomposition}
Technical details on both data reduction techniques are discussed in the following appendices to keep the manuscript self-contained. We apply the POD in the form of the method of snapshots~\citep{Sirovich1987,Bailon2011} on the vector $\mathbf{g} =(u_x', u_z', b'')^T$, such that its $k^{\rm th}$ component can be written as
%------------------------------------------------------
\begin{equation}
    \label{eq:pod}
    g_k(x,z,t) = \sum\limits_{i = 1}^{N_{\rm dof}} a_i(t) \Phi_i^{\rm (k)}(x,z).
\end{equation}
%------------------------------------------------------
This linear method decomposes the scalar field $g_k$ into time dependent coefficients $a_i(t)$ and spatial modes $\Phi_i^{\rm (k)}(x,z)$, such that the truncation error is minimized. The degrees of freedom $N_{\rm dof}$ can then be reduced, by taking only $N_{\rm POD} \ll N_{\rm dof}$ modes and coefficients with the most variance into account.
%------------------------------------------------------
\begin{equation}
    \label{eq:pod_rom}
    g_k(x,z,t) \approx \sum\limits_{i=1}^{N_{\rm POD}} a_i(t) \Phi_i^{\rm (k)}(x,z). %\qquad N_{\rm POD} \ll N_{\rm dof}.
\end{equation}
%------------------------------------------------------
As mentioned above, we consider the $N_{\rm POD} = 300$ most energetic POD time coefficients as the input for the ESN. The cumulative contributions of the first $N_{\rm POD} = 300$ POD modes for the three cases of $\beta=0.1$, $0.2$, and $0.3$ capture then 82.4\%, 80.2\%, and 78.4\% of the original total energy (kinetic energy plus temperature variance), respectively. 

%------------------------------------------------------
\begin{figure}[!htbp]
    \centering
    \includegraphics[width = .4\textwidth]{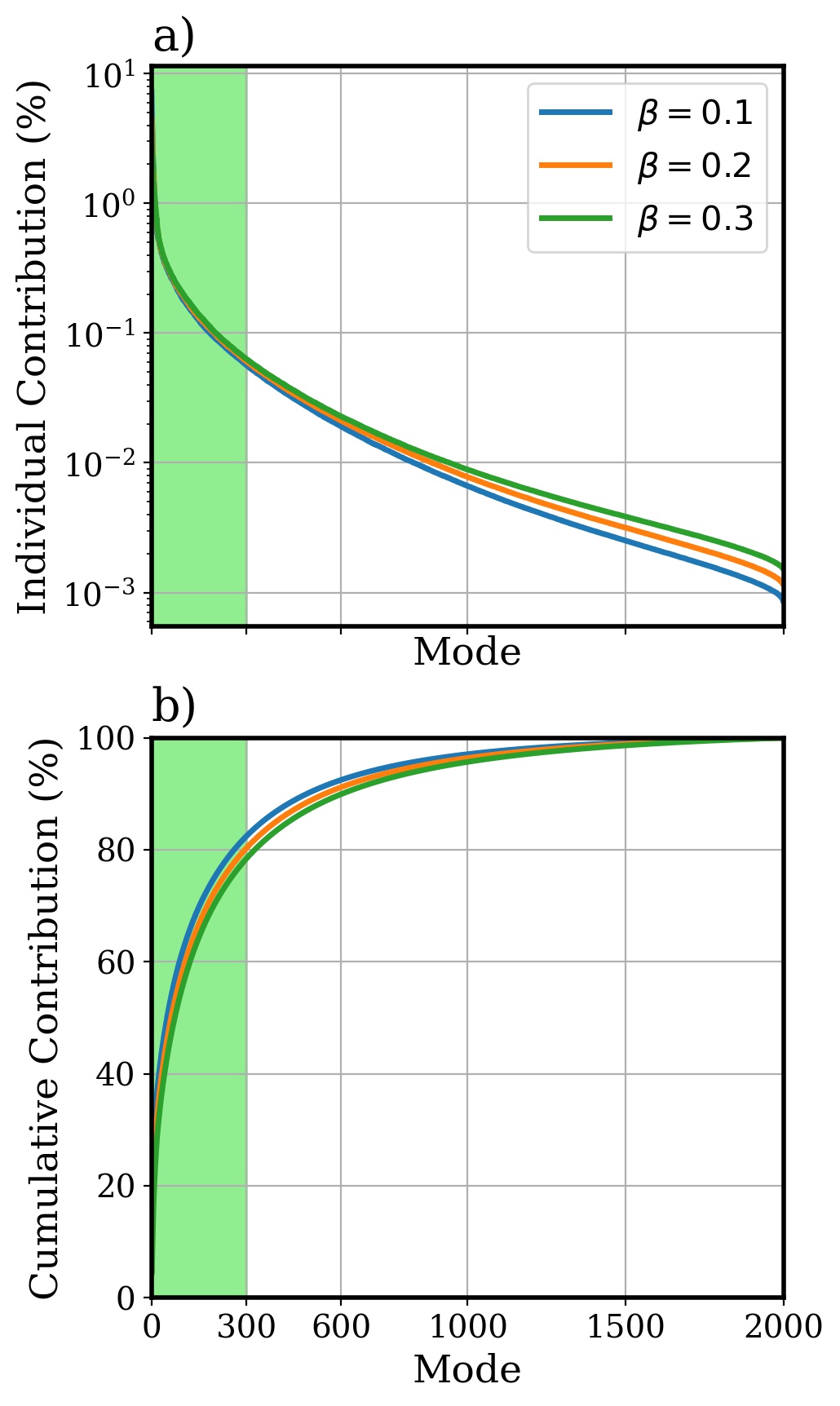}
    \caption{Spectrum of eigenvalues of the POD modes. a) Individual contribution of each POD mode to the total energy. b) Cumulative contribution. The green shaded area marks the contribution of the first 300 modes.}
    \label{fig:pod_spectrum}
\end{figure}
%------------------------------------------------------

\section{Convolutional autoencoder}
\label{subsubsec:conv_autoencoder}
An autoencoder is a feed-forward neural network which is trained to reproduce its network input $\mathbf{g}$ as network output~\cite{Baldi2012,Goodfellow-et-al-2016}. In order for the network to not just copy its inputs to the output layer, an intermediate bottleneck structure is introduced, such that the original information is compressed to an encoding or latent low-dimensional space. Therefore, the autoencoder consists of two parts which are trained as one network. The encoder $\mathbf{f}$ compresses the high-dimensional inputs to a low-dimensional representation
\begin{align}
    \label{eq:encoder}
    \mathbf{\xi} = \mathbf{f}_{\theta_{\rm encoder}}(\mathbf{g})
\end{align}
where $\mathbf{\xi} \in \mathbb{R}^{N_{CAE}}$ is the encoding or latent space and $\theta_{\rm encoder}$ includes all trainable weights and biases of the encoder network.

%------------------------------------------------------
\begin{table}[!hptb]
    \renewcommand{\arraystretch}{1.5} 
    \centering
    \begin{tabular}{lccc}
         \hline\hline
         $\beta$  & $0.1$ & $0.2$ & $0.3$ \\
         \hline
         Training Loss ($\%$)& $6.2\times 10^{-4}$ & $7.1\times 10^{-4}$ & $8.1\times 10^{-4}$\\
         Validation Loss ($\%$)& $2.66\times 10^{-3}$ & $2.76\times 10^{-3}$ & $2.72\times 10^{-3}$\\
         \hline\hline
    \end{tabular}
    \caption{Mean squared error loss of the CAE reconstruction after 1200 epochs of training. Each CAE was trained on $8000$ snapshots of $(u_x',u_z',b'')$ for their corresponding $\beta$. The validation loss was computed on $2000$ different snapshots.}
    \label{tab:cae_loss}
\end{table}
%------------------------------------------------------
\begin{table*}[!htbp]
\renewcommand{\arraystretch}{1.5} 
\centering
\begin{tabular}{cccccccccc}
     \hline\hline
     Layer & Conv\#1 & Conv\#2 & Conv\#3 &Conv\#4 &Conv\#5 &Conv\#6 &Conv\#7 &Conv\#8 &Conv\#9  \\
     \hline
     channels & (3,8) & (8,16) & (16,16) & (16,32) & (32,16) & (16,16) & (16,8) & (8,3) & (3,3)\\
     kernel & (7,7) & (5,5) & (3,3) & (3,3)& (3,3) & (3,3) & (3,3) & (5,5) &(7,7)\\
     MP kernel & (2,1) & (2,2) &(2,2)& (2,2) & (2,2) & (2,2) & (2,2) & (2,1) & -\\
     \hline\hline
\end{tabular}
\caption{Size of convolutional channels, kernel and max pooling kernel for each layer in the autoencoder network. The channels are given in the form (input channel, output channel), while both the convolutional and MP kernel are given by (height, width). The shape of the input data was (3,30,720).}
\label{tab:appendix_cae}
\end{table*}
%------------------------------------------------------

The decoder $\mathbf{h}$ then attempts to decode the encoded latent space and reconstruct the original information
\begin{align}
    \label{eq:decoder}
    \mathbf{g}^{\rm AE} = \mathbf{h}_{\theta_{\rm decoder}}(\mathbf{\xi}) 
    =\mathbf{h}_{\theta_{\rm decoder}}\left(\mathbf{f}_{\theta_{\rm encoder}}(\mathbf{g})\right).
\end{align}
Here $\mathbf{g}^{\rm AE}$ is the autoencoder reconstruction and $\theta_{\rm decoder}$ includes all trainable weights and biases of the decoder network. We use a convolutional autoencoder (CAE) which makes use of convolutional layers that have proven to be extremely useful in pattern detection and classification of images~\cite{Krizhevsky2012}. While the $\mathbf{\xi}$ can be understood as a low-dimensional representation of the input, similar to the POD time coefficients $\mathbf{a}$, the trained weights and biases correspond to the POD spatial modes which contain information on how to decode the latent space. The training of the CAE requires backpropagation of errors through the convolutional networks. An optimally working CAE minimizes the difference between original input and final output, $\mathbf{g}^{\rm AE}\approx \mathbf{g}$.

As for the POD approach, we take snapshots of $\mathbf{g} = (u_x', u_z', b'')^T$ of $\beta = 0.1 - 0.3$ as input for their own CAE. Finally one can use the trained encoder to translate the flow dynamics into dynamics of the latent space $\mathbf{\xi}(t)$. We choose an encoding dimension of $N_{CAE}=300$ and train the network with $8000$ snapshots of $\mathbf{g}$ and use $2000$ further snapshots to validate its performance. The training and validation mean square error loss of each CAE is listed in table \ref{tab:cae_loss}. Out of the $2000$, to the CAE unseen snapshots, we sample the $700$  time steps of $\xi^{\beta=0.1}(t)$, used for training the ENS, and $700$ time steps of $\xi^{\beta=0.2}(t)$ and $\xi^{\beta=0.3}(t)$, used for validation of the ESN predictions.

We use a CAE with four convolutional layers and one dense layer in the encoder and five convolutional layers and one dense layer in the decoder. Except the last layer in the decoder, each convolutional layer is complemented by a Max-Pooling (MP) operation, in order to downsample the input data. Further, all layers are followed by a batch normalization and dropout layer. We find that batch normalization stabilizes the training process and dropout reduces the effect of overfitting, where the neural network shows poor performance on the validation data. The activation function of the last layer in both encoder and decoder was sigmoid, while all other layers were followed by a \textit{Parametric Rectified Linear Unit} (PReLU)~\cite{He2015}. The channel size, as well as convolutional and max pooling kernels are listed in table \ref{tab:appendix_cae}. Using this architecture the total number of trainable weights and biases amounts to $4.96\times 10^6$.

The autoencoder is trained using the \textit{ADAM} optimizer~\cite{Diederik2017} with a learning rate $10^{-5}$, batch size $64$ and a L$_2$-norm penalty term with penalty parameter $10^{-6}$. The loss function that was minimized was chosen to be the mean square error between input and output fields. Moreover, the input data was scaled to the range $\left[0,1\right]$ before it was passed to the input layer of the CAE. Finally, the network was trained for 1200 epochs on 2 GPUs and took about $106$min.

\section{Echo state network grid search procedure}
%------------------------------------------------------
\begin{table}[!htbp]
\renewcommand{\arraystretch}{1.5} 
\centering
\begin{tabular}{lcccc}
    \hline\hline
     &$N_{\rm}$, & $D$, & $\gamma$ & $\varrho$\\
     \hline
     range & $\left[512,4096\right]$ & $\left[0.1,1.0\right]$ & $\left[0.1,1.0\right]$ &$\left[0.1,2.0\right]$\\
     no. samples& $4$& $100$& $10$& $100$ \\
     \hline\hline
\end{tabular}
\caption{ESN grid search range of each hyperparamter that was studied. The number of samples indicates how many different values of each hyperparameter were studied.}
\label{tab:appendix_rcm}
\end{table}
%------------------------------------------------------
In order to find an optimal reservoir for both reduction methods and both $\beta$ values, we conducted grid searches on four important reservoir hyperparameters, namely $N_{\rm}$, $D$, $\gamma$ and $\varrho$ when training the case of $\beta=0.1$. The range and number of different values of each hyperparameter study are listed in table \ref{tab:appendix_rcm}.

\subsection{Results for $\beta = 0.3$}
\begin{figure*}[!htbp]
	\centering
	\includegraphics[width = .95\textwidth]{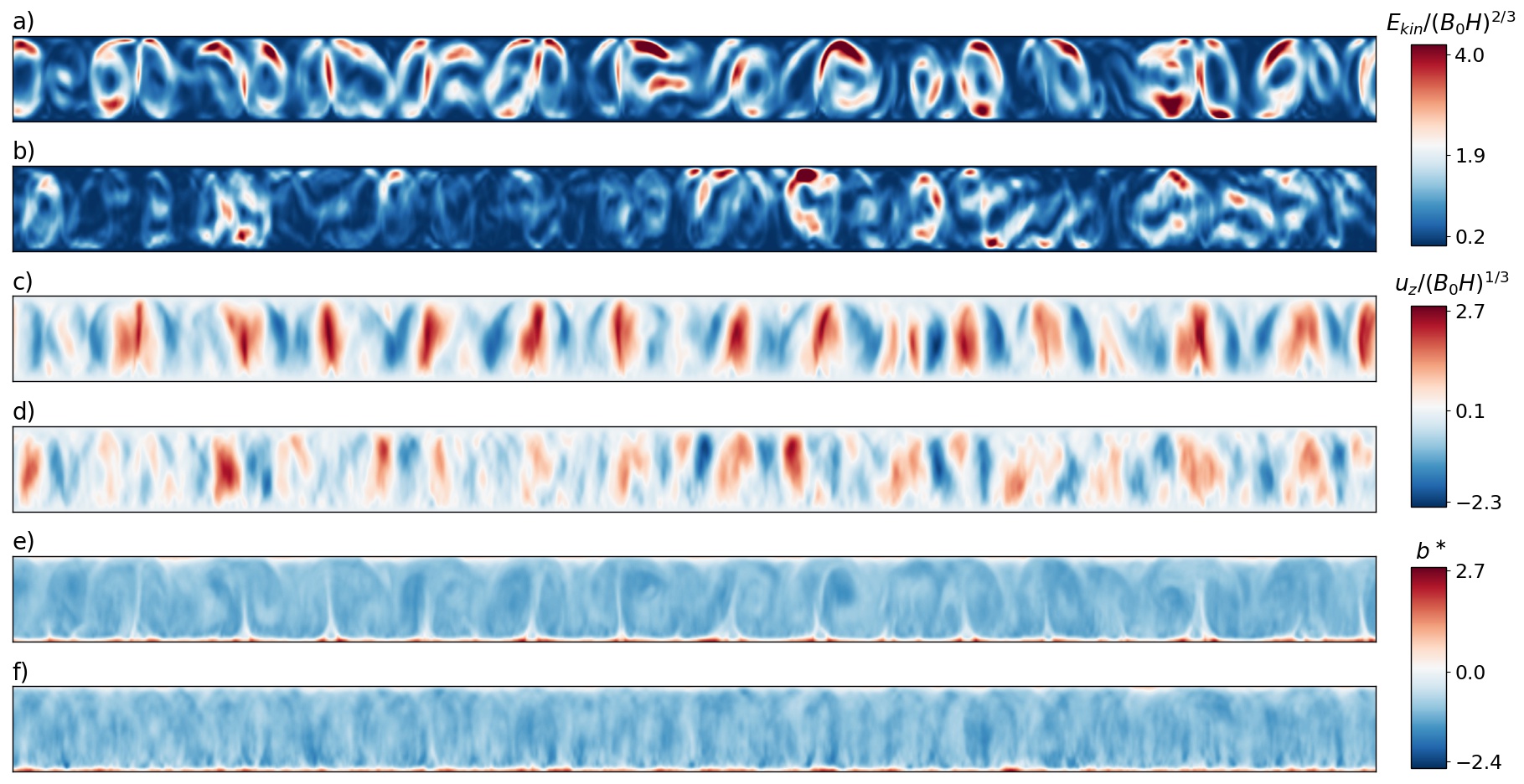}
	\caption{POD case for inferring $\beta = 0.3$. Instantaneous snapshots of the local turbulent kinetic energy $E_{\rm kin}(x,y,t_0)$ in panels (a,b), the vertical velocity component $u_z$ in panels (c,d) and the normalized buoyancy $b^\ast$ in panels (e,f) at time step $n=350$ in the prediction phase. POD reconstructions with the most energetic $N_{\rm POD}$ modes of $\beta = 0.2$ (validation snapshot) are shown in panels (a), (c), and (e). The corresponding ESN predictions are displayed in panels (b), (d), and (f).}
	\label{fig:app1}
\end{figure*}

\begin{figure*}[!htbp]
	\centering
	\includegraphics[width = .95\textwidth]{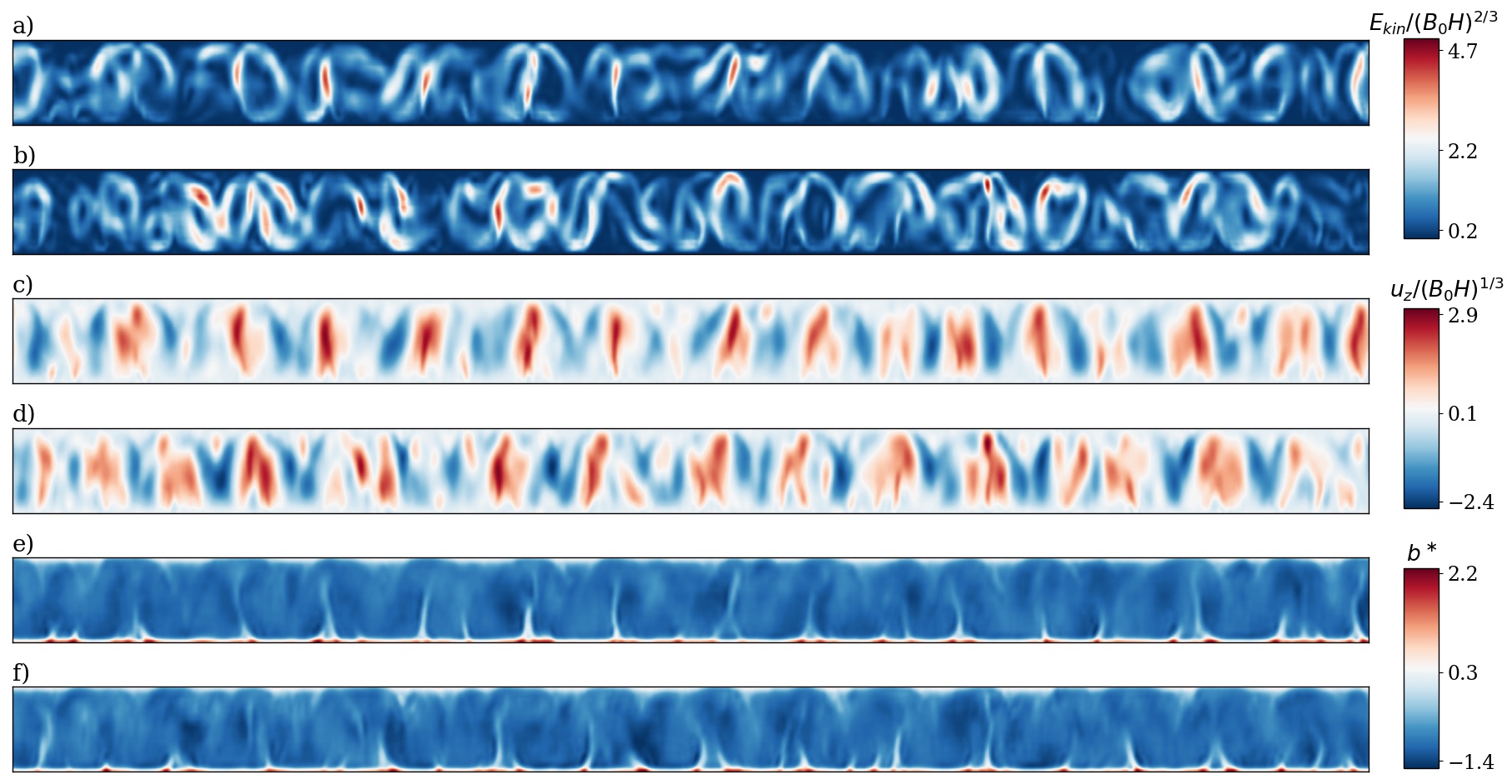}
	\caption{CAE case for inferring $\beta = 0.3$. Instantaneous snapshots of the turbulent kinetic energy $E_{\rm kin}(x,y,t_0)$ in panels (a,b), the vertical velocity component $u_z$ in panels (c,d) and the normalized buoyancy $b^\ast$ in panels (e,f) at time step $n=350$ in the prediction phase. CAE reconstructions of $\beta = 0.3$ (validation snapshot) are shown in panels (a), (c), and (e). The corresponding ESN predictions are displayed in panels (b), (d), and (f).}
	\label{fig:app2}
\end{figure*}

For completeness, we show in figures \ref{fig:app1} and \ref{fig:app2} the results for $\beta=0.3$ that correspond to those for the case $\beta=0.2$ in the main text. See figures 8 and 10. 
%------------------------------------------------------
%------------------------------------------------------
%------------------------------------------------------
\clearpage
%% ------------------------------
%% References and Citations
%% ------------------------------

%\bibliography{HeyMelSchu}

\end{document}